\documentclass[runningheads]{llncs}
\usepackage[T1]{fontenc}
\usepackage{graphicx}
\usepackage[colorlinks=true,linkcolor=blue,citecolor=blue]{hyperref}%

\usepackage{color}

\usepackage[font=small,skip=2pt,labelfont=bf]{caption}

\usepackage{subcaption}

\usepackage{enumerate}

\usepackage{algorithmic}
\usepackage{amsmath}

\begin{document}
\title{Exploring Encrypted Keyboards to Defeat Client-Side Scanning in End-to-End Encryption Systems}
%

\titlerunning{Exploring Encrypted Keyboards to Defeat Client-Side Scanning}
%
\author{Mashari Alatawi \and Nitesh Saxena}
\authorrunning{Mashari Alatawi and Nitesh Saxena}
%
\institute{Texas A\&M University, College Station TX 77843, USA\\
\email{\{mashari,nsaxena\}@tamu.edu}}
\maketitle              
\begin{abstract}
End-to-End Encryption (E2EE) aims to make all messages impossible to read by anyone except you and your intended recipient(s). Many well-known and widely used Instant-Messaging (IM) applications (such as Signal, WhatsApp, Apple’s iMessage, and Telegram) claim to provide an E2EE functionality. However, a recent technique called client-side scanning (CSS), which could be implemented by these IM applications, makes these E2EE claims grandiose and hollow promises. The CSS is a technology that scans all sending and receiving messages from one end to the other, including text, images, audio, and video files. Some in industry and government now advocate this CSS technology to combat the growth of malicious child pornography, terrorism, and other illicit communication. Even though combating the spread of illegal and morally objectionable content is a laudable effort, it may open further backdoors that impact the user's privacy and security. Therefore, it is not end-to-end encryption when there are censorship mechanisms and backdoors in end-to-end encrypted applications. In this paper, we shed light on this hugely problematic issue by introducing an encrypted keyboard that works as a system keyboard and can be enabled on the user's phone device as a default system keyboard. Therefore, it works on every application on the user's phone device when the user is asked to enter some data. To avoid the CSS system, users can use this encrypted keyboard to encrypt and decrypt their messages locally on their phone devices when sending and receiving them via IM applications. We first design and implement our encrypted keyboard as a custom keyboard application, and then we evaluate the effectiveness and security of our encrypted keyboard. Our study results show that our encrypted keyboard can successfully encrypt and decrypt all sending and receiving messages through IM applications, and therefore, it can successfully defeat the CSS technology in end-to-end encrypted systems. We also show that our encrypted keyboard can be used to add another layer of E2EE functionality on top of the existing E2EE functionality implemented by many end-to-end encrypted applications.

\end{abstract}

\vspace{-1mm}
\section{Introduction}
Smartphones are becoming increasingly important in today's society as more people rely on them for daily communication. Due to this widespread use of smartphones, many applications are continuously being developed to meet people's needs and facilitate their text, audio, and video communications. However, people today are concerned about the security and privacy of their communications due to government surveillance programs and law enforcement agencies, which have pushed them to worry about their online activities and sharing sensitive information. As a consequence of these concerns, many Instant-Messaging (IM) applications have been developed to address these issues by providing secure messaging solutions through a method known as End-to-End Encryption (E2EE) which protects conversations from any third party. However, these IM applications, which claim to provide an E2EE feature, are vulnerable to Man-in-the-Middle (MitM) attacks, either by compromising the service providers or by using another form of attack.

Furthermore, the E2EE functionality has been plagued by a recent attack called \textit{client-side scanning}, \textit{endpoint filtering}, or \textit{local processing}, which breaks the E2EE feature claimed by IM applications \cite{erica2019}. In this client-side scanning (CSS) system, an end-to-end encrypted application performs a scan against any text, image, audio, or video in the message before encrypting the message and sending it to the intended recipient. If the CSS system in the end-to-end encrypted application finds any matching item, it will prevent the user from sending the message or reporting any matching item to government censorship or law enforcement authorities. It is a laudable effort when end-to-end encrypted applications tend to use the CSS technique to prevent child exploitation imagery (CEI), thwart terrorism, or provide copyright protection. However, this will lead to open another door for further censorship mechanisms and build further backdoors that impact the user's privacy. Having such CSS technology in end-to-end encrypted applications could be abused by many attackers. Thus, it might be causing more threats to the user's privacy than protecting objectionable content. People are now worried that the Meta company is listening in on their WhatsApp conversations, which are supposed to be encrypted from end to end, to show them ads that are more relevant to them \cite{askhn2022ads}. Recently, the Apple company has also proposed its CSS system to fight child sexual abuse materials (CSAMs) over the Internet \cite{paul2021}. It uses a database of known CSAM image hashes maintained by child safety organizations and reports any matching image to law enforcement agencies. This database is nothing but a set of hashes whereby each image is converted into a different unique numeric representation. A hash function, which is a computer function that maps data of arbitrary size to fixed-size values called hash values, converts such an image into a small hash value, and only that hash value is converted into that image. Also, it has come out that the Federal Bureau of Investigation (FBI) and its international partners secretly ran an encrypted messaging app called \textit{Anom} to spy on and collect tens of millions of messages from \textit{Anom} users \cite{cox2022}. Their goal was to monitor organized crime on a global scale by looking over the shoulders of organized criminals as they talked to each other. The revealed parts of the code showed that the exchanged messages were secretly duplicated and sent to a third party (which is the FBI and its law enforcement partners) that was hidden from the users’ contact lists. Therefore, from a security perspective, that is not an E2EE functionality when having these censorship mechanisms and backdoors in end-to-end encrypted applications. The end-to-end encrypted applications should provide the E2EE feature in such a way that the exchanging of messages is known only to the sender and the intended recipient. No third party, not even the service provider, should be privy to any message content exchanged between the sender and the intended recipient. 

In this paper, we introduce an encrypted keyboard to address this issue facing the E2EE functionality implemented by IM applications. This encrypted keyboard is a system keyboard that the user can enable on his phone device and therefore use to encrypt or decrypt a message. Many IM applications implement the CSS system and still advertise that they provide the E2EE feature. They may argue that the CSS mechanism will just occur right before and after the encryption and decryption of messages while keeping the promise of providing the E2EE functionality to take place between two endpoints. However, this will be a hollow promise, and there will not be an E2EE anymore since the service provider is sitting on both endpoints and watching over the user's shoulder to filter all sending and receiving messages. The goal of our encrypted keyboard is to encrypt multimedia data (including text, image, audio and video) locally on the user's phone before the user puts them into any IM app (like WhatsApp) and then decrypt them when they reach the other end. Our encrypted keyboard on the user's phone device will ensure that any CSS system implemented by an IM application will be prevented. Our approach not only protects against CSS technologies but also strengthens the E2EE feature used by current end-to-end encrypted applications by implementing it twice.

\textbf{Contributions:} Our contributions are as follows:
\vspace{-1mm}
\begin{itemize}
	\item \textbf{Encrypted Keyboard for Preventing Client-Side Scanning:} We introduce our encrypted keyboard, built as a system keyboard, that can effectively prevent CSS technologies in many end-to-end encrypted systems. Our encrypted keyboard can be enabled on the user's phone device as a primary keyboard; therefore, it works on every application on the user's phone device that requires inputs from the user. We believe that our encrypted keyboard can provide a great solution to secure users' messages from filtering techniques and other surveillance mechanisms that technology companies may use.
	\item \textbf{Design and Implementation of the Encrypted Keyboard:} We design an encrypted keyboard that follows a similar layout to one of the most popular keyboard layouts, such as a QWERTY English keyboard. Our implementation consists of a custom keyboard application that allows users to encrypt and decrypt their data locally on their phone devices. It also allows users to display the decrypted data on the custom keyboard's interface. This custom keyboard application can be installed on the users' phones like any other application. Users can then enable this custom keyboard as the default system keyboard on their phone devices and, therefore, can use it to encrypt data locally on their phone devices before entering them into IM applications. It also works locally on users' phones to decrypt data that were encrypted and sent to them through IM apps.
	\item \textbf{Evaluating the Encrypted Keyboard for Effectiveness and Security:} We evaluate the effectiveness of our encrypted keyboard by testing its ability to encrypt and decrypt the user's data. Here we focus only on the ability to defeat such a CSS system by encrypting and decrypting the user’s data locally on his phone device. We show that our encrypted keyboard can not only allow users to encrypt their data locally on their phone devices but is also able to decrypt their encrypted data locally on their phone devices. We also establish that our encrypted keyboard can encrypt and decrypt the user's data before or after exchanging them through IM applications. Our results show that our encrypted keyboard can be effective against CSS technologies by encrypting and decrypting users' data locally on their phone devices. Our encrypted keyboard may also be used to enhance the security of E2EE functionality implemented by many end-to-end encrypted applications by adding an extra layer of E2EE functionality on top of their E2EE functionality.
\end{itemize}

\vspace{-1mm}
\section{Background}
Secure messaging applications aim to provide private communications in such a way that sensitive information is hidden from anyone who is not a part of these communications. This can be done through an E2EE functionality, which ensures that all private messages are only viewable by the sender and the intended recipient. Due to the Snowden revelations about widespread government surveillance in 2013, people were concerned about their security and privacy in online communications \cite{mary2014}. Therefore, IM and Voice over IP (VoIP) applications began integrating E2EE security features to make communication more secure. Several IM and VoIP applications, including WhatsApp \cite{whatsapp}, Telegram \cite{telegram}, Signal \cite{signal}, Viber \cite{viber}, and Skype \cite{skype}, have adopted the E2EE protocol to protect all private communications in recent years. Although many IM and VoIP applications claim to implement the E2EE protocol to secure private communications, they vary in their goals, ambiguous security claims, threat models, usability, and adoption properties \cite{unger2015sok,rottermanner2015privacy}.

Furthermore, several end-to-end encrypted applications may implement the CSS technology to scan messages just before they are sent from a sender or after they are received by an intended recipient. Using this technology, end-to-end encrypted applications aim to scan and flag any message before transmission, thereby preventing the transmission of any message or item that may contravene legal prohibitions. This could be done by adding a scanning system as a part of the end-to-end encrypted application; in other words, the scanning mechanism could be built into the end-to-end encrypted application such as Signal, WhatsApp, and many others \cite{rosenzweig2020law}. These applications may use related software to check such a message against a database of problematic content (such as CSAM images), extremist content, copyright infringement, rumors, or misinformation. This means that if any match has been found, it may block the message or report it to law enforcement authorities. From their perspective, these end-to-end encrypted applications claim that adding such a CSS system can protect against any illegal and morally objectionable content. However, even though this point of view is considered a well-intentioned attempt to help ban such content, it could open many doors to expanding the scope of the CSS system \cite{abelson2021bugs}. Backdoors and censorship in end-to-end encrypted applications will break the E2EE feature, even though these applications promise and guarantee their users that their messages will be encrypted between two endpoints and that the CSS system will only scan their messages right before encrypting them or right after decrypting them \cite{erica2019}. The E2EE feature should ensure that a message is only seen by the sender and the recipient. This means that no one else can read the message or scan its content to figure out what it is about.

\vspace{-1mm}
\subsection{Related Work}
There has been a vast amount of work studying secure messaging solutions. Borisov et al. \cite{borisov2004off} proposed a protocol, called “off-the-record messaging”, in 2004 for secure online communication. Their protocol was designed to provide perfect forward secrecy and deniable authentication for messages exchanged between two users. It was inspired by the notion of having a private conversation in a room between two people, Alice and Bob. In this scenario, Alice will be confident that no one else outside the room can hear the private conversation between her and Bob. Also, she will be confident that no one, not even Bob, can go to court and blame her by using her words against her. The OTR protocol has been plugged into different IM applications such as Pidgin \cite{pidgin}; however, it has not been adopted widely due to its usability shortcomings \cite{stedman2008user}. Frosch et al. \cite{frosch2016secure} analyzed the Signal protocol to provide a detailed analysis of its underlying cryptographic protocol and highlight its claimed security features. Similarly, Cohn-Gordon et al. \cite{cohn2020formal} performed a formal security analysis of the Signal protocol as a multi-stage authenticated key exchange protocol. They showed some standard security properties and showed that these properties meet the protocol's security claims. Further, Unger et al. \cite{unger2015sok} conducted a comprehensive academic survey on secure communication tools in terms of investigating their security, usability, and ease of adoption properties. They systematized these tools and discussed three fundamental issues of secure messaging solutions: trust establishment, conversation security, and transport privacy.

On the other hand, there is prior work that has studied the CSS technology in terms of the security and privacy issues of using this technology. Abelson et al. \cite{abelson2021bugs} studied the potential security and privacy risks of utilizing CSS technologies. They argued that these systems could be exploited to open many doors that may impact the privacy and security of communications, even though the initial objective of these systems is solely to prevent the spread of illegal and morally objectionable content. Then, people would struggle to stop the system's expansion and prevent its abuse. Another study by Reis et al. \cite{reis2020can} showed that it is possible to scan messages on end-to-end encrypted systems. They explored the idea of using fact-checking to detect misinformation in WhatsApp and proposed an architecture that could be implemented by WhatsApp to detect and flag misinformation on users' devices. This might introduce new system vulnerabilities and effectively violate E2EE's privacy and security guarantees. To address this, we propose an encrypted keyboard that protects end users’ devices from CSS technologies, which could be implemented by numerous end-to-end encrypted applications.

There are some encrypted keyboard applications available on the Apple App Store and the Google Play Store, such as Enigma Encryption Keyboard \cite{enigma} and WhisperKeyboard \cite{whisperkeyboard}, that aim to provide end-to-end encryption and decryption for text messages. While these applications focus only on encrypting and decrypting text messages, we consider encrypting and decrypting not only text messages but also other multimedia messages (like images, audio, and video) in our encrypted keyboard. We also consider using an automated process to decipher encrypted text messages in our encrypted keyboard, whereas these applications require a user to copy the encrypted text message to the phone’s clipboard in order to decipher the encrypted text message, thereby adding extra burden to users every time they want to decode their encrypted text messages.

\vspace{-1mm}
\subsection{Threat Model}
As a reminder, the goal of the E2EE functionality is to protect the contents of a message against anyone who is not involved in the private conversation. Therefore, we assume the same threat model as outlined in a comprehensive survey on secure messaging by Unger et al. \cite{unger2015sok}. The authors stated that the threat model includes the following attackers:

\begin{itemize}
	\item \textbf{Local Adversary:} An (active/passive) attacker who can control local networks on any side of the conversation, such as owners of open wireless access points.
	\item \textbf{Global Adversary:} An (active/passive) attacker who can control many parts of the Internet service (e.g., powerful nation-states or large internet service providers).
	\item \textbf{Service Providers:} All service operators could be considered potential attackers when IM and VoIP applications utilize a centralized infrastructure for distributing public keys and storing or forwarding messages (such as a public-key directory).
\end{itemize}

However, in this work, we extend our threat model to include the endpoints of end-to-end encrypted applications, since many of these applications could use CSS technology. We assume that the CSS technology is made as a part of an end-to-end encrypted application such as Signal or WhatsApp, which means that it is built into the end-to-end encrypted application. It is possible that these applications perform pre-scanning of the users' messages by sitting on any endpoint and looking over the user's shoulder when he sends a message. Therefore, these applications should not be trusted by users and should be considered potential adversaries. We assume that the operating system of a phone device is healthy and secure at both endpoints. We also assume that the attacker does not have a deeper access to information on a device. In other words, the attacker only has direct access to a database and files that are associated with the end-to-end encrypted application (e.g., WhatsApp) installed on the user's phone device but does not have internal control of the other parts of the user's phone system, including other applications' data on the user's phone device.

\vspace{-1mm}
\section{Design and Implementation}
\begin{figure}[t]
	\centering
	\includegraphics[width=0.75\textwidth]{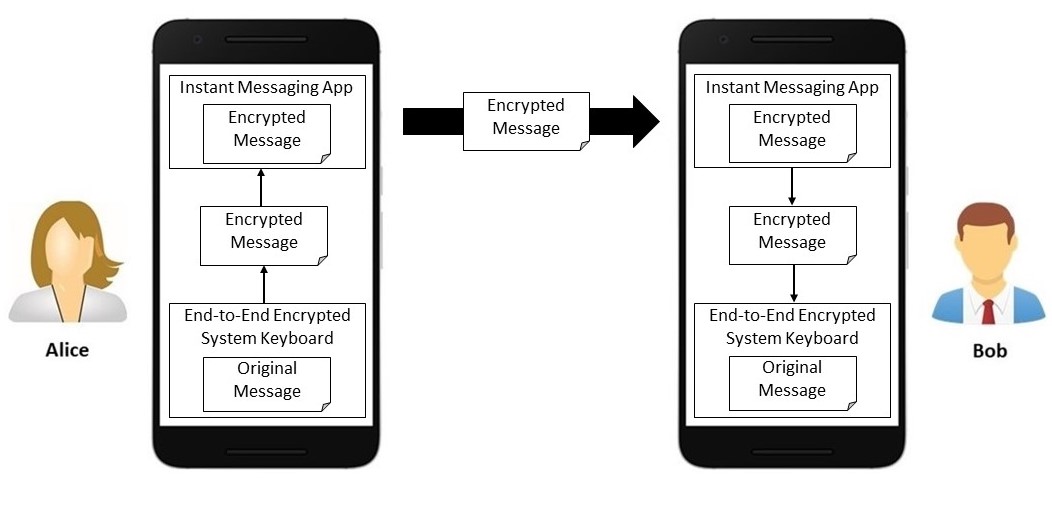}
	\caption{The architecture of our encrypted keyboard. The sender (Alice) creates a message, encrypts it using the custom keyboard, and then sends it via an instant messaging app. At the other end, the receiver (Bob) gets the encrypted message in the instant messaging app and uses the custom keyboard to decrypt it.} \label{fig1}
	\vspace{-1mm}
\end{figure}

We designed a system keyboard application using Android Studio to demonstrate the ability to provide a protection mechanism against CSAM image-related techniques, copyright infringement, or any material whose illegality is uncontested. In addition to the CSS technology, our encrypted keyboard aims to provide an extra layer of E2EE security over the existing E2EE functionality implemented by the end-to-end encrypted applications. Our encrypted keyboard architecture is shown in Figure \ref{fig1}.

\vspace{-1mm}
\subsection{End-to-End Encrypted System Keyboard}
\begin{figure*}[h]
	\centering
	\begin{subfigure}{.19\textwidth}
		\centering
		\includegraphics[width=1\linewidth]{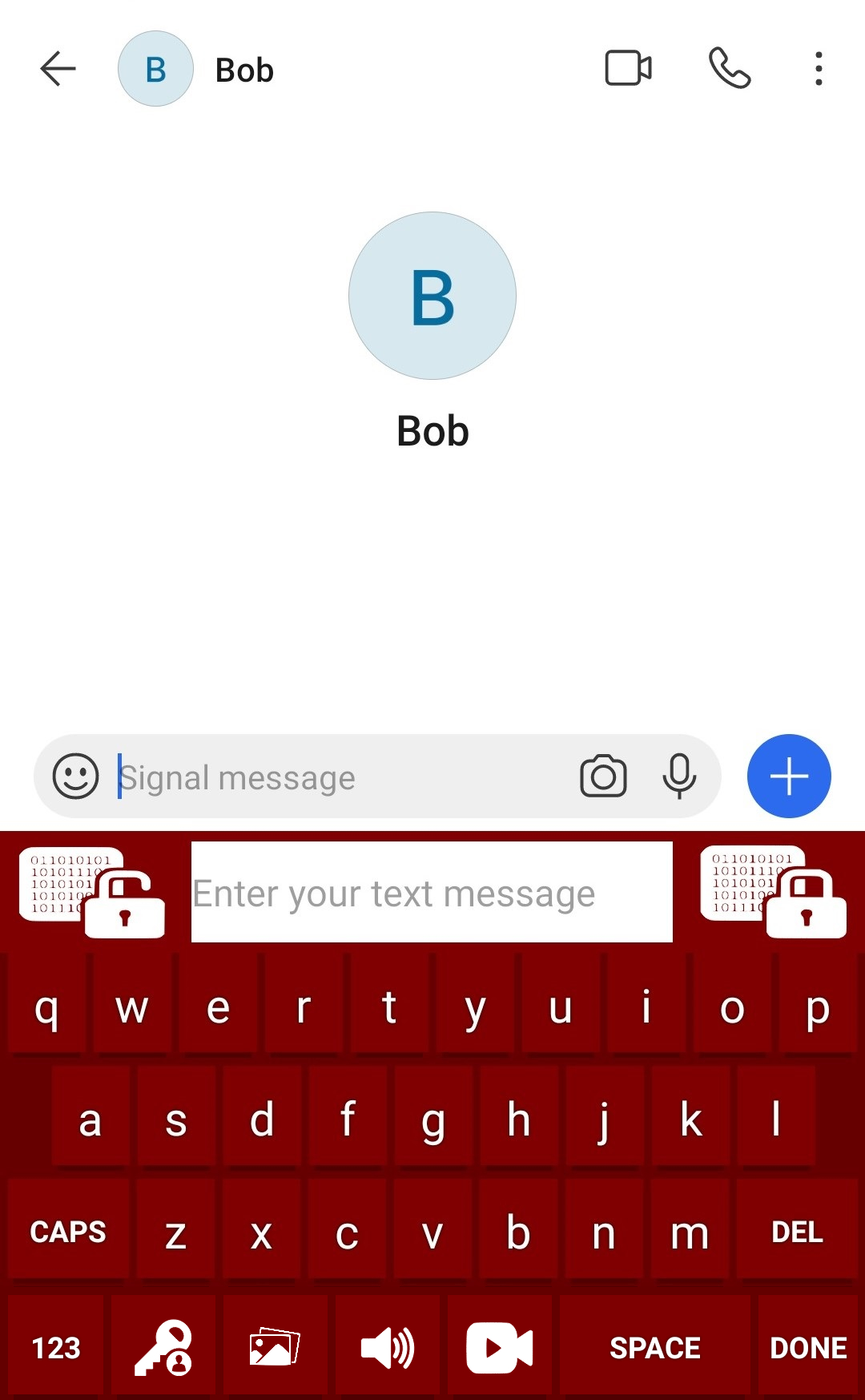}
		\caption{QWERTY keyboard layout}
		\label{fig2:subfig1}
	\end{subfigure}\hspace{1mm}%
	\begin{subfigure}{.19\textwidth}
		\centering
		\includegraphics[width=1\linewidth]{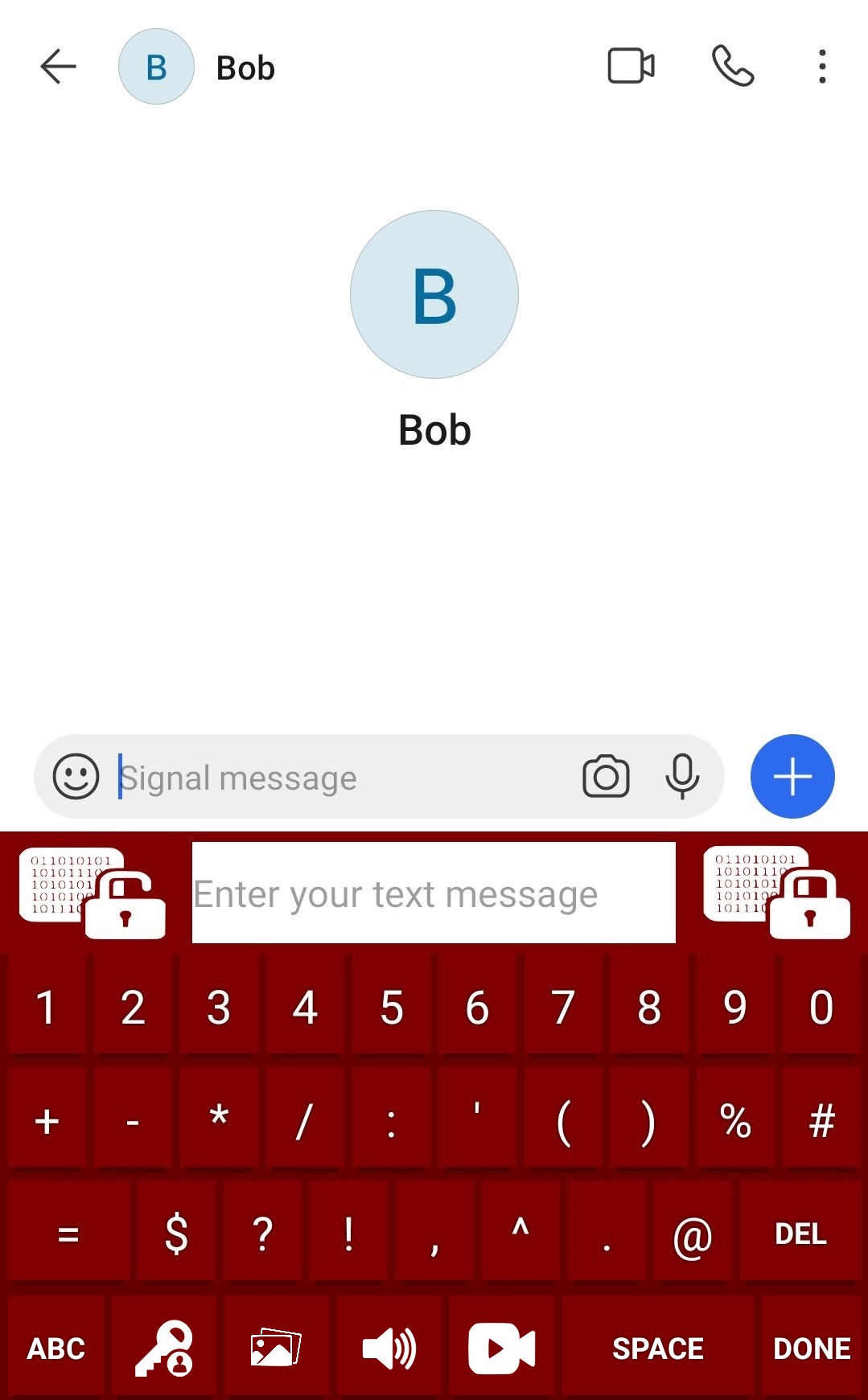}
		\caption{Numeric keyboard layout}
		\label{fig2:subfig2}
	\end{subfigure}\hspace{1mm}%
	\begin{subfigure}{.19\textwidth}
		\centering
		\includegraphics[width=1\linewidth]{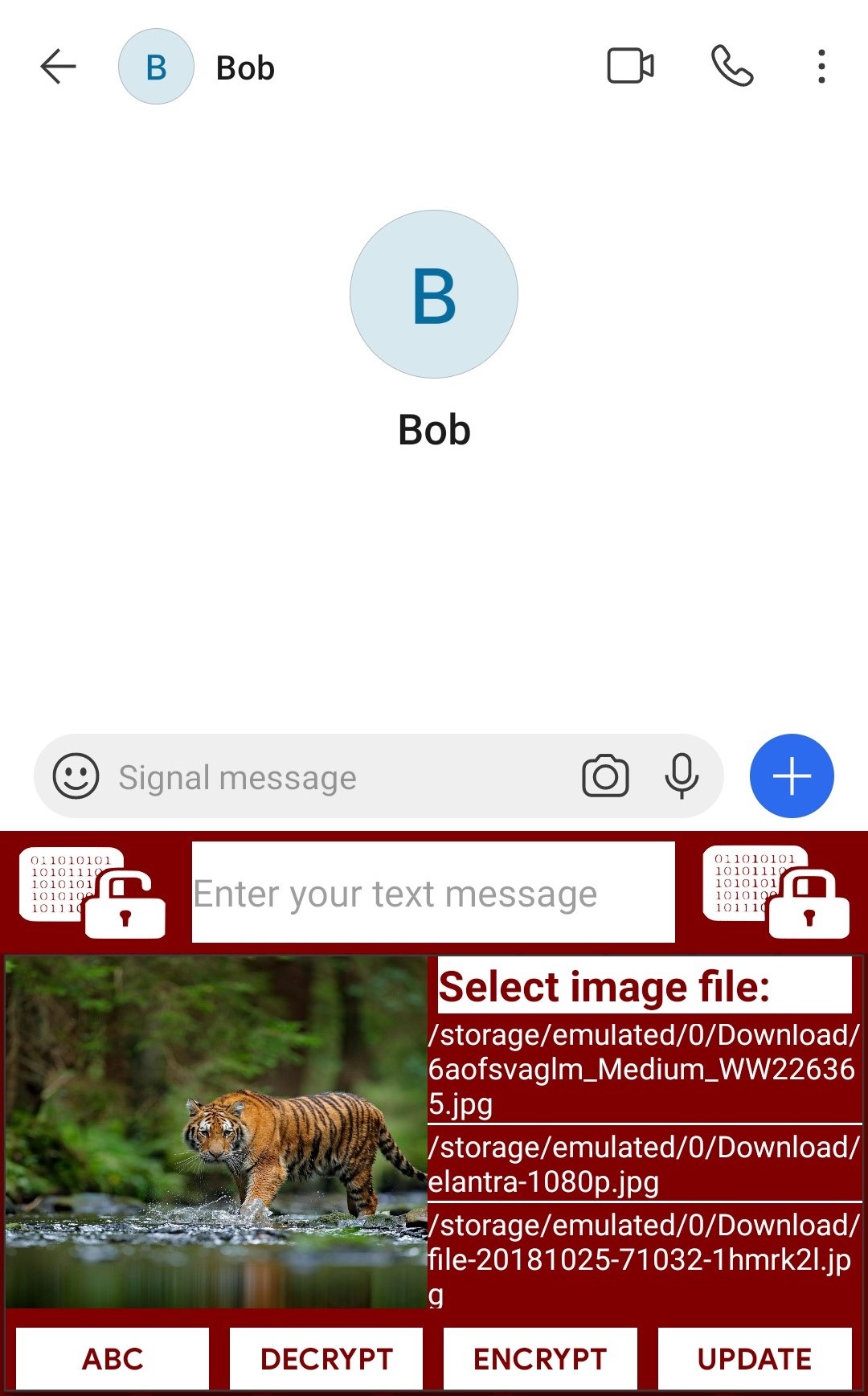}
		\caption{Image keyboard layout}
		\label{fig2:subfig3}
	\end{subfigure}\hspace{1mm}%
	\begin{subfigure}{.19\textwidth}
		\centering
		\includegraphics[width=1\linewidth]{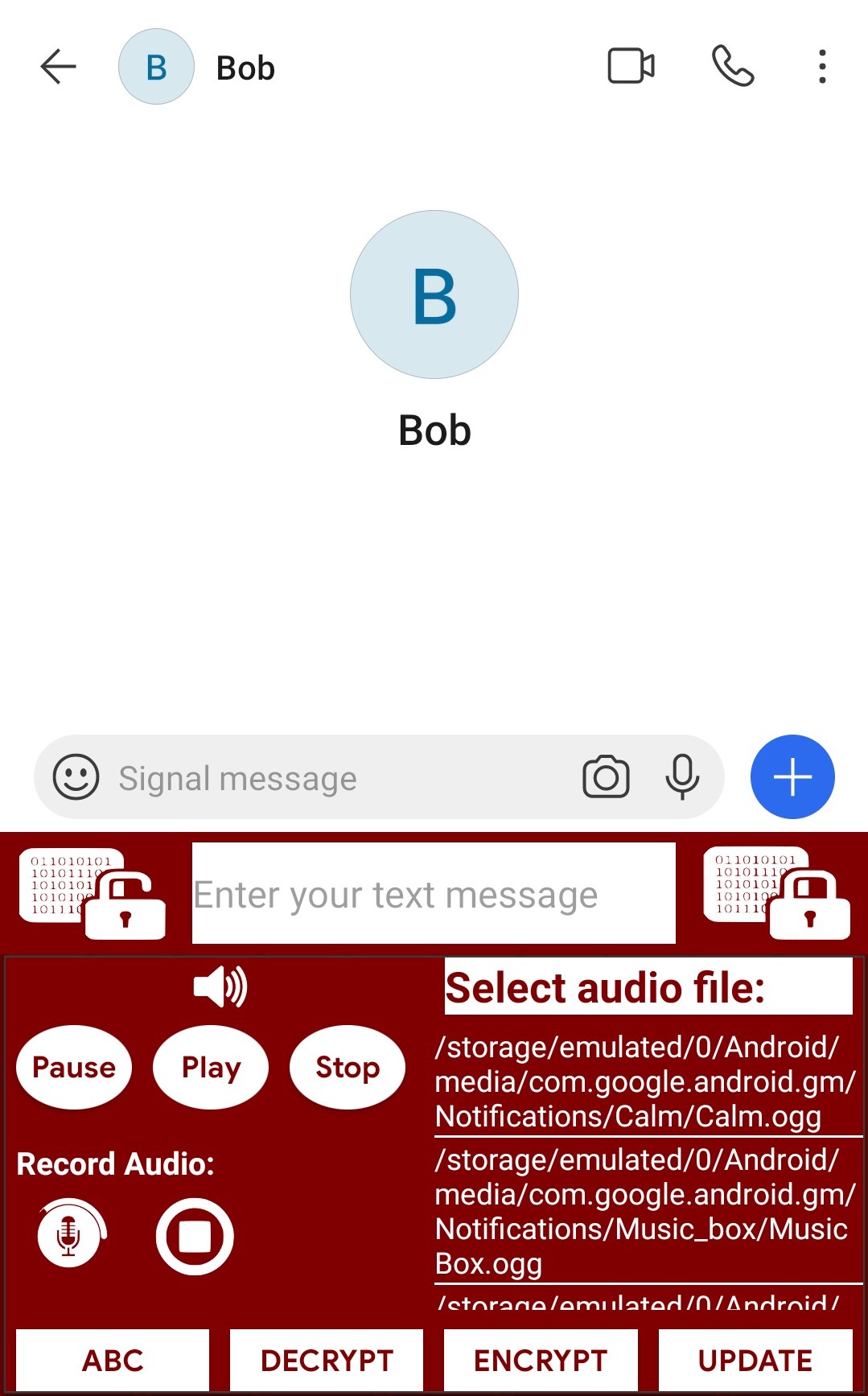}
		\caption{Audio keyboard layout}
		\label{fig2:subfig4}
	\end{subfigure}\hspace{1mm}%
	\begin{subfigure}{.20\textwidth}
		\centering
		\includegraphics[width=1\linewidth]{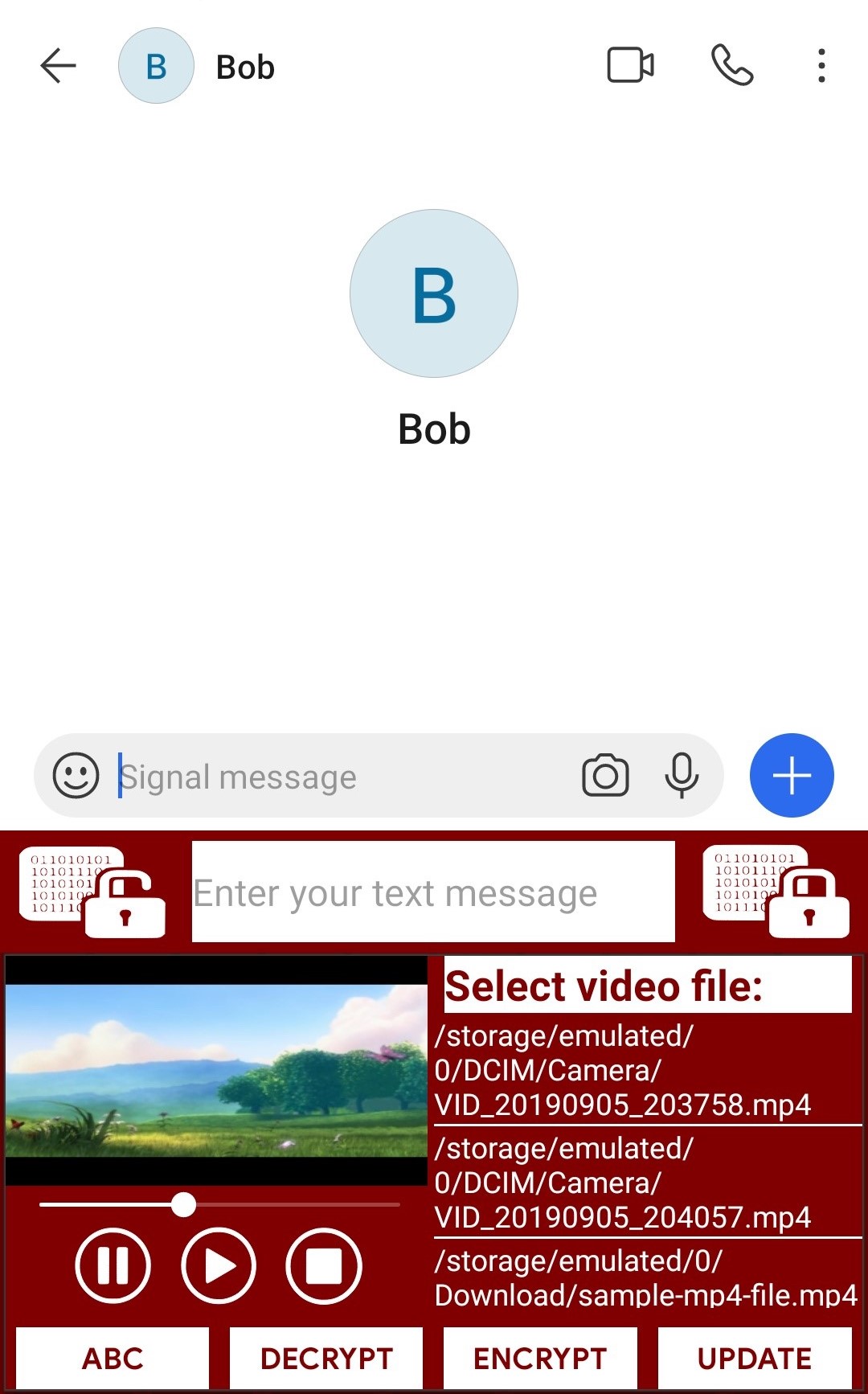}
		\caption{Video keyboard layout}
		\label{fig2:subfig5}
	\end{subfigure}
	\caption{User interface of our encrypted keyboard}
	\label{fig2:user interface}
	\vspace{-1mm}
\end{figure*}

We developed an Android application that implements the input method service needed by an Input Method Editor (IME) to get connected to the Android system. Therefore, our encrypted keyboard runs as a system keyboard and creates an input method that allows the user to enter encrypted text into any IM application. In addition to encrypting text, our encrypted keyboard allows the user to encrypt images, audio, and video files locally on his phone device before sending them via any IM application. Another key design component of our encrypted keyboard is to build it as an end-to-end encrypted system, thereby adding another layer of E2EE security on top of the current E2EE functionality implemented by many end-to-end encrypted applications. In order to enable our encrypted keyboard on the user’s phone device, the user needs to install our encrypted keyboard application on his phone device and then navigate to the \textit{Language and input} setting in his phone’s system settings, where he can select this encrypted keyboard to be the default keyboard on his phone device. The user can then use this encrypted keyboard in any application that requires data entry. The user interface of our encrypted keyboard is depicted in Figure \ref{fig2:user interface}.

We designed five different interfaces for our encrypted keyboard. First, we designed a user interface layout to encrypt/decrypt text and followed a similar layout for one of the most popular keyboard layouts, such as a standard QWERTY English keyboard with an additional numeric keyboard layout (see Figures \ref{fig2:subfig1} and \ref{fig2:subfig2}). This user interface layout contains additional elements (i.e., encryption and decryption buttons, an edit text field, etc.) to implement the E2EE functionality. It also contains keys at the bottom of the interface to allow the user to navigate between other user interfaces. Then, we designed an additional three interfaces for encrypting and decrypting other multimedia elements such as images, audio, and video (see Figures \ref{fig2:subfig3}, \ref{fig2:subfig4}, and \ref{fig2:subfig5}). In each user interface, there is a list that contains all the existing user's images, audio, or video files on his phone device. The user can select any image, audio, or video file from the list and display that image on the image box, play the audio file using audio control buttons, or play the video file using video control buttons on the current keyboard interface. Moreover, in the audio keyboard layout, we added audio recording buttons to allow the user to record voice memos. We put the encryption and decryption buttons at the bottom of these new user interfaces so that the user can encrypt and decrypt his multimedia data on his phone.

\vspace{-1mm}
\subsection{Strong Encryption Algorithm}
In order to provide confidentiality to the data, we used the Advanced Encryption Standard (AES) to encrypt and decrypt multimedia data. It is one of the most powerful encryption algorithms that is widely used in many different technology fields and computer systems. It is also one of the most efficient encryption algorithms, and it is considered to be fast and flexible, thereby ensuring the security of data and making it trustworthy for users \cite{abdullah2017advanced}. AES is one of the symmetric encryption algorithms where both the sender and the receiver have the same key to encrypt and decrypt data. It uses three different key sizes, such as 128, 192, and 256 bits, with a 128-bit block cipher. The details of the encryption and decryption processes of the AES algorithm can be found in \cite{abdullah2017advanced,nist2001specification,daemen2002advanced}. We adopted an open-source library for implementing the encrypted text task \cite{encryptionLibrary}. It creates an encryption instance with the AES algorithm in Cipher Block Chaining (CBC) mode and uses a key size of 128 bits. The same key (the secret key) was used for the encryption and decryption processes of all text messages between two parties. We designed on our encrypted keyboard at the bottom of the main user interface, as shown in Figure \ref{fig5:secret key interface}, a key button to allow the user to select the shared secret key with his intended recipient. We discuss the management and distribution of the keys that provide secure communications functionality in Section \ref{keysmanagement}.

\begin{figure*}[h]
	\centering
	\begin{subfigure}{.22\textwidth}
		\centering
		\includegraphics[width=1\linewidth]{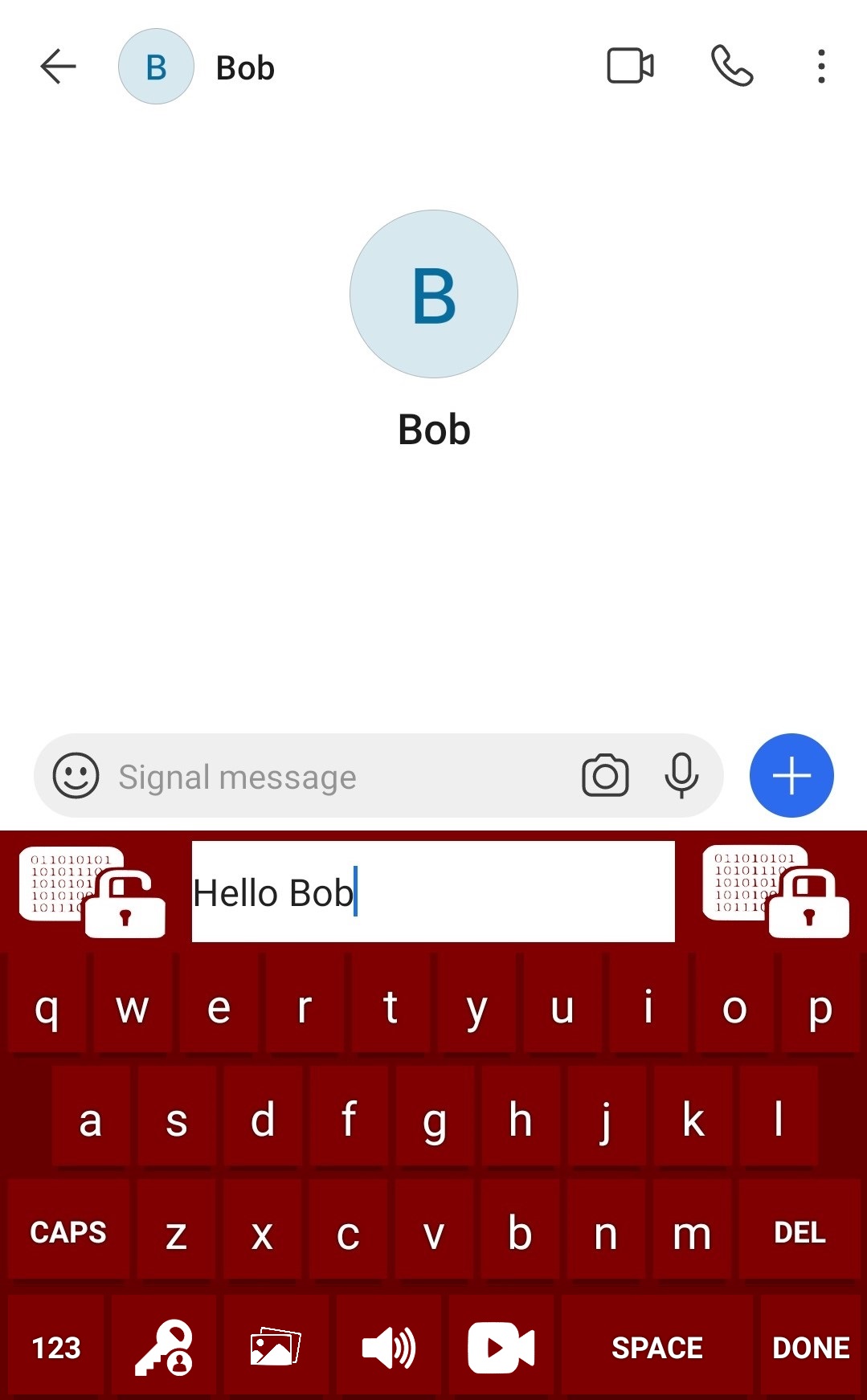}
		\caption{Alice composes a text for Bob using the private text box.}
		\label{fig3:subfig1}
	\end{subfigure}\hspace{3mm}%
	\begin{subfigure}{.22\textwidth}
		\centering
		\includegraphics[width=1\linewidth]{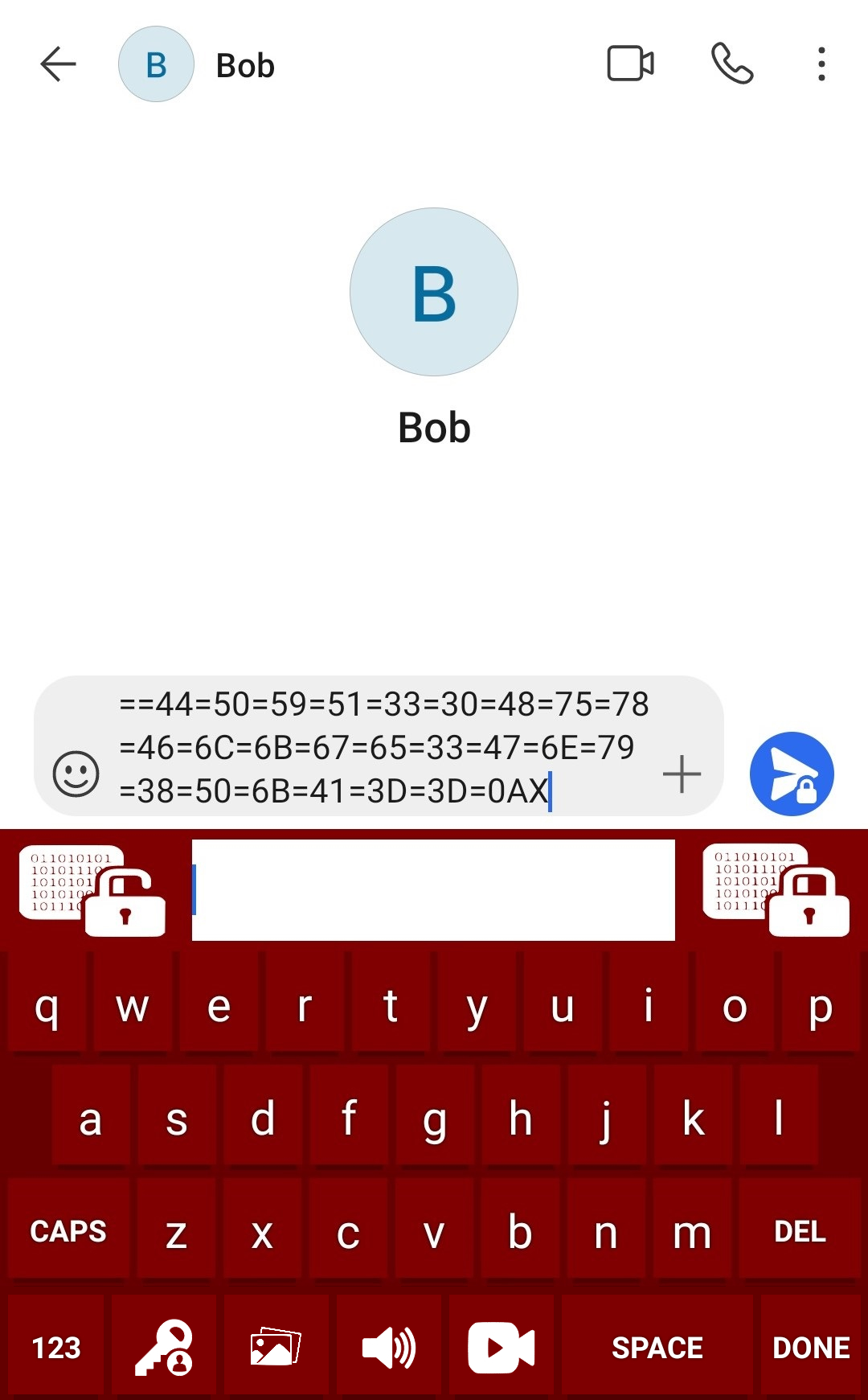}
		\caption{Alice encrypts her text message to Bob by tapping the encryption key located at the top right corner of the encrypted keyboard.}
		\label{fig3:subfig2}
	\end{subfigure}\hspace{3mm}%
	\begin{subfigure}{.22\textwidth}
		\centering
		\includegraphics[width=1\linewidth]{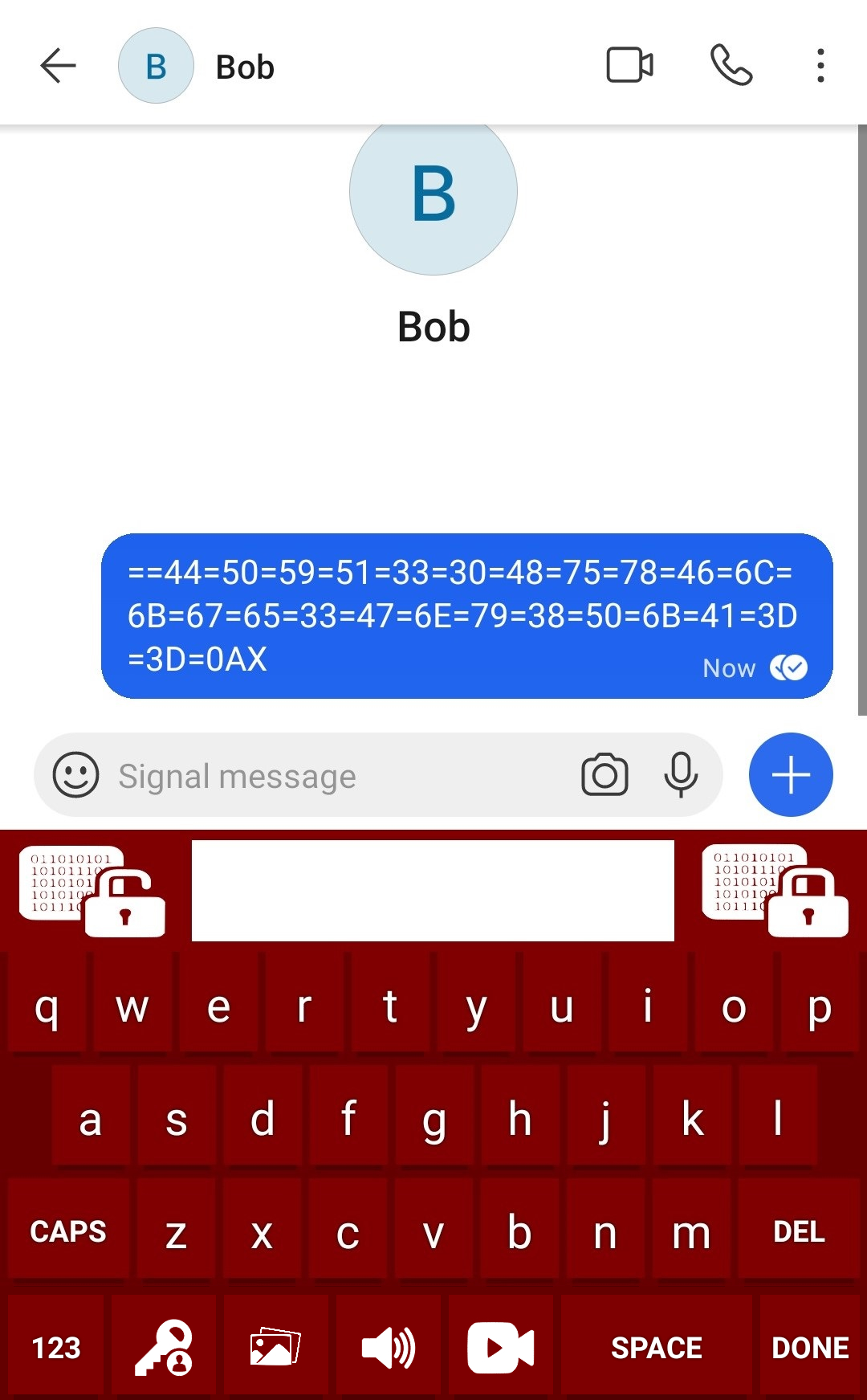}
		\caption{Alice sends her encrypted text message to Bob.}
		\label{fig3:subfig3}
	\end{subfigure}\hspace{3mm}%
	\begin{subfigure}{.22\textwidth}
		\centering
		\includegraphics[width=1\linewidth]{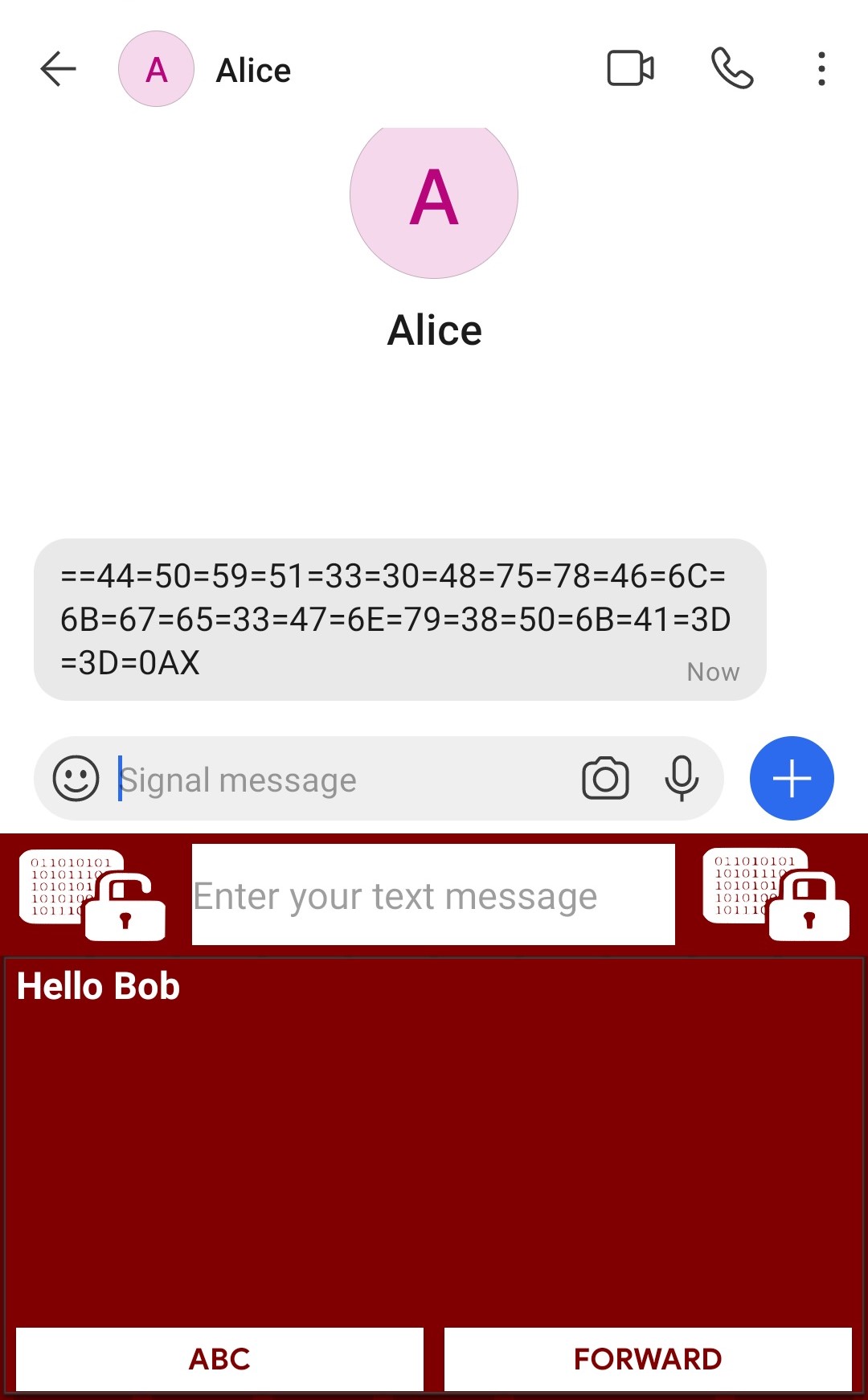}
		\caption{Bob receives the encrypted text message from Alice and decrypts it by tapping the decryption key located at the top left corner of the encrypted keyboard.}
		\label{fig3:subfig4}
	\end{subfigure}
	\caption{Encryption and decryption of the text message using our encrypted keyboard when the Signal application is used to send and receive the encrypted text message exchanged between Alice and Bob.}
	\label{fig3:user interface encrypt/decrypt text}
	\vspace{-1mm}
\end{figure*}

In order to secure all the user’s textual messages and to prevent any IM application from performing the CSS technique on the user’s phone device, we used a local edit text (private editor) as a private text box where the user can use it for typing and modifying the text. Furthermore, we created a new form of interaction between this private editor and our encrypted keyboard using (the InputConnection interface in Java code) in order to receive all typed text on the encrypted keyboard. Once the user enters the text into the private text box and clicks the (encryption button), our encrypted keyboard encrypts the currently composing text, which is the text located inside the private text box, before entering the encrypted text into the text field linked to the currently open IM application. After that, the user can send only the encrypted text to the other end via the current IM application (e.g., Signal) (see Figures \ref{fig3:subfig1}, \ref{fig3:subfig2}, and \ref{fig3:subfig3}). Once the receiver receives the encrypted text message, he can decrypt it by tapping the (decryption button) to obtain the original text message that was sent by the sender (see Figure \ref{fig3:subfig4}).

The basic idea underlying using our encrypted keyboard is to allow users to encrypt their multimedia data locally on their phone devices, thereby avoiding any implementation of CSS technologies. Furthermore, we implement an E2EE feature for messages sent from one party of the private conversation to the other. This E2EE feature could be used to enhance the security against attackers listening onto the channel when users exchange their messages over an end-to-end encrypted application (e.g., WhatsApp) since users can encrypt their multimedia data using our encrypted keyboard before inserting them into the end-to-end encrypted application, which will then encrypt them once again using its E2EE functionality. Our encrypted keyboard can guarantee that no messages will be revealed to those attackers even if they can somehow compromise the encrypted messages transmitted by any end-to-end encrypted application. Therefore, when our encrypted keyboard is used in an IM application that also has an E2EE feature, it will add an extra layer of E2EE security.

\vspace{-1mm}
\subsection{Automated Decryption Process}
In our study, we considered the usability aspect in terms of automating the decryption process in order to make it easier to decrypt any encrypted text on the user's phone device. We aim to reduce the burden on the user when he unscrambles encrypted text messages on his phone device. SMASheD \cite{mohamed2016smashed} utilizes the Android Debug Bridge (ADB) functionality to have access to phone device resources/services. Therefore, we adapted the SMASheD server and pushed it to the phone device to run a screenshot service every second in the background and to store the screenshot image in a file. Our implementation does not store all these screenshot images. Instead, this image file will be overwritten every time a screenshot image is taken, thereby avoiding the consumption of too many resources. Even though many real-world applications can take screenshots and are used daily, our encrypted keyboard is a more trustworthy application because it works as a system keyboard and does not send any images outside the user's phone device during the actual implementation. The screenshot service will keep running in the background until the user decides to stop it or the phone device is switched off. Our goal in running this screenshot service on a phone device is to get the content displayed in the foreground of a phone device's screen to use as an input file for an Optical Character Recognition (OCR) engine. The OCR engine is used to turn a screenshot into text that can be edited and searched \cite{shinde2012text}. The encrypted text is then taken from the output of the OCR engine so that our decryption method can get the needed ciphertext and the decryption phase can be completed.

Thus, if the encrypted text message is received by the receiver's phone device, we expect that a screenshot image of the current phone device's screen has already been taken by the running screenshot service before the user clicks on the (decryption button). Reading and extracting text from the screenshot image are then needed to obtain the ciphertext from the image. To this end, we adopted the (tess-two) project to run the Tesseract OCR engine on the screenshot image. Tesseract, which was developed at HP between 1984 and 1994, is an open-source OCR engine. It was adjusted and enhanced in 1995 for better accuracy before HP released it as open source in late 2005 \cite{smith2007overview}, which is now available at \cite{tesseractocr}. Once our decryption method in the decryption phase gets the ciphertext, it will automatically decipher the ciphertext and return the original text that was sent by the sender.

Because of the accuracy of OCR tools, which ranges from 71\% to 98\% \cite{patel2012optical}, we convert an encrypted text into a hexadecimal format, where numbers are represented by a base of 16, before sending it to the other end user. The purpose of converting the encrypted text into hexadecimal format is to increase the accuracy of our OCR performance by limiting the OCR engine to recognizing a small group of characters. Another reason for recognizing a small group of characters is to avoid any noisy and garbled results that may occur if the OCR engine reads all possible characters. Therefore, we set up a white list for our OCR engine that contains hexadecimal symbols from 0 to 9, corresponding to number values from 0 to 9, and A to F, corresponding to number values from 10 to 15. By converting an encrypted text into a hexadecimal format, we believe that any accuracy issue that might result from our OCR performance can be surmounted.

\vspace{-1mm}
\subsection{Multimedia Support}
\begin{figure*}[h]
	\centering
	\begin{subfigure}{.22\textwidth}
		\centering
		\includegraphics[width=1\linewidth]{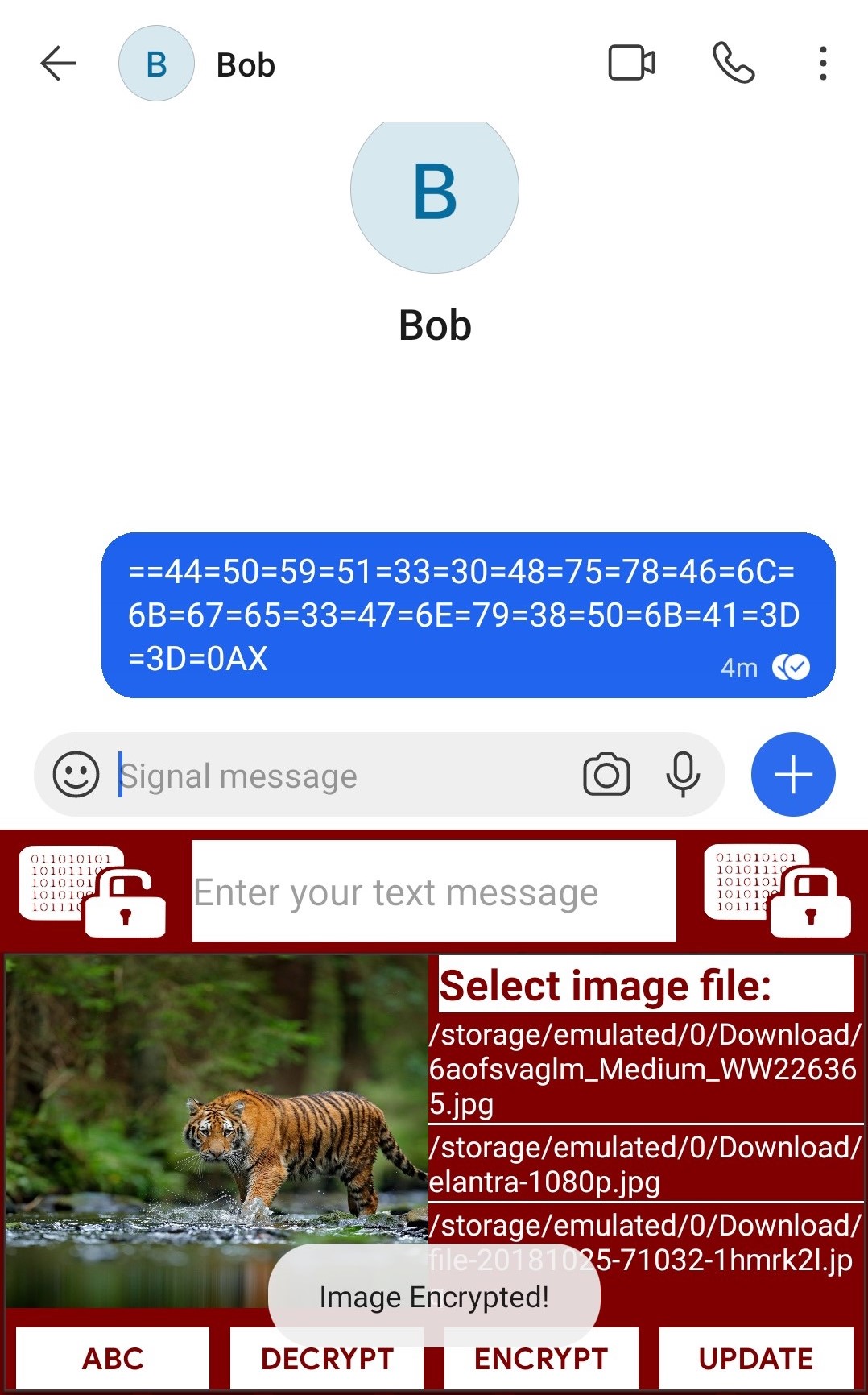}
		\caption{Alice encrypts the image file locally on her phone device by tapping the ENCRYPT button located at the bottom of the encrypted keyboard.}
		\label{fig4:subfig1}
	\end{subfigure}\hspace{3mm}%
	\begin{subfigure}{.22\textwidth}
		\centering
		\includegraphics[width=1\linewidth]{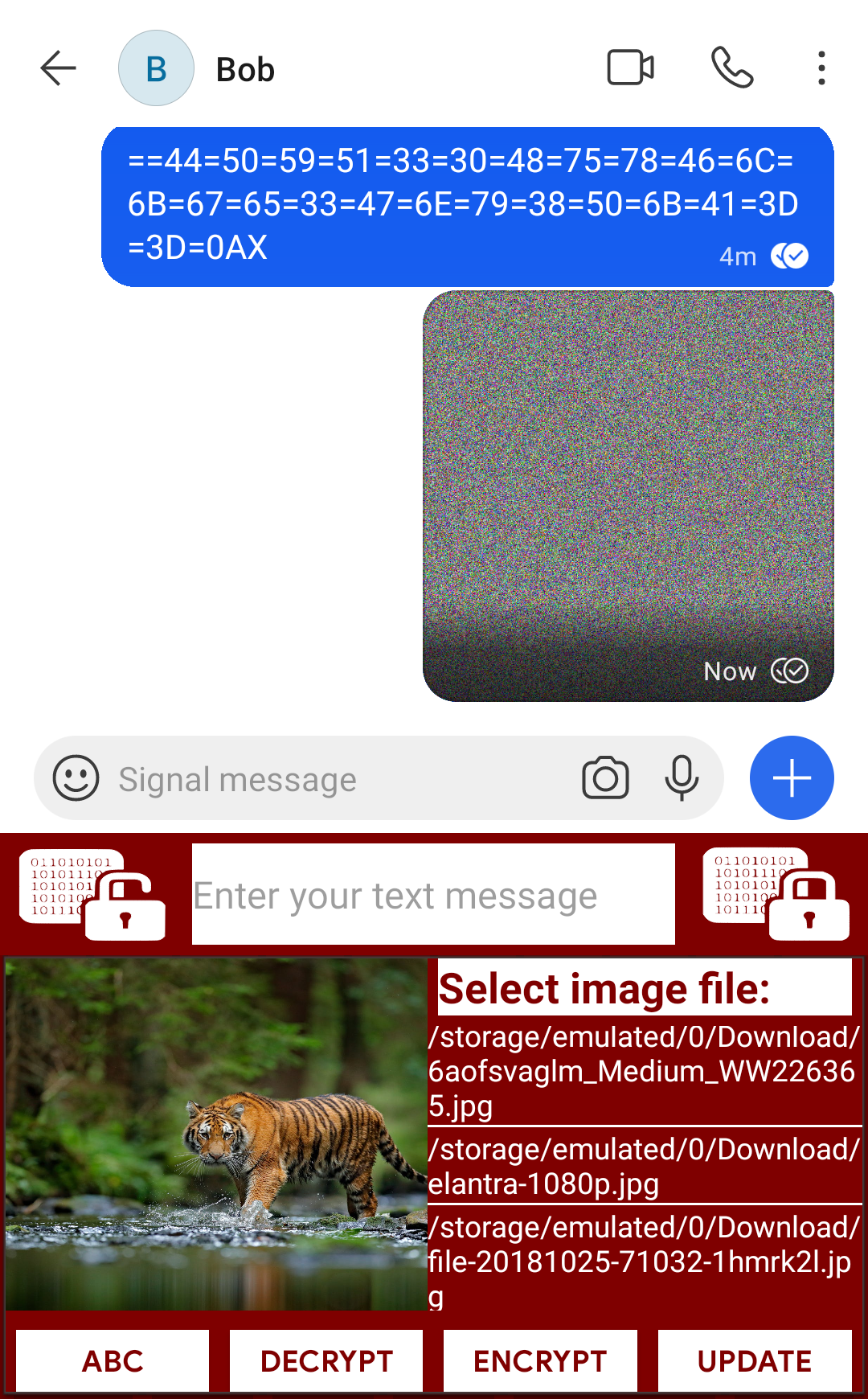}
		\caption{Alice sends the encrypted image file to Bob.}
		\label{fig4:subfig2}
	\end{subfigure}\hspace{3mm}%
	\begin{subfigure}{.22\textwidth}
		\centering
		\includegraphics[width=1\linewidth]{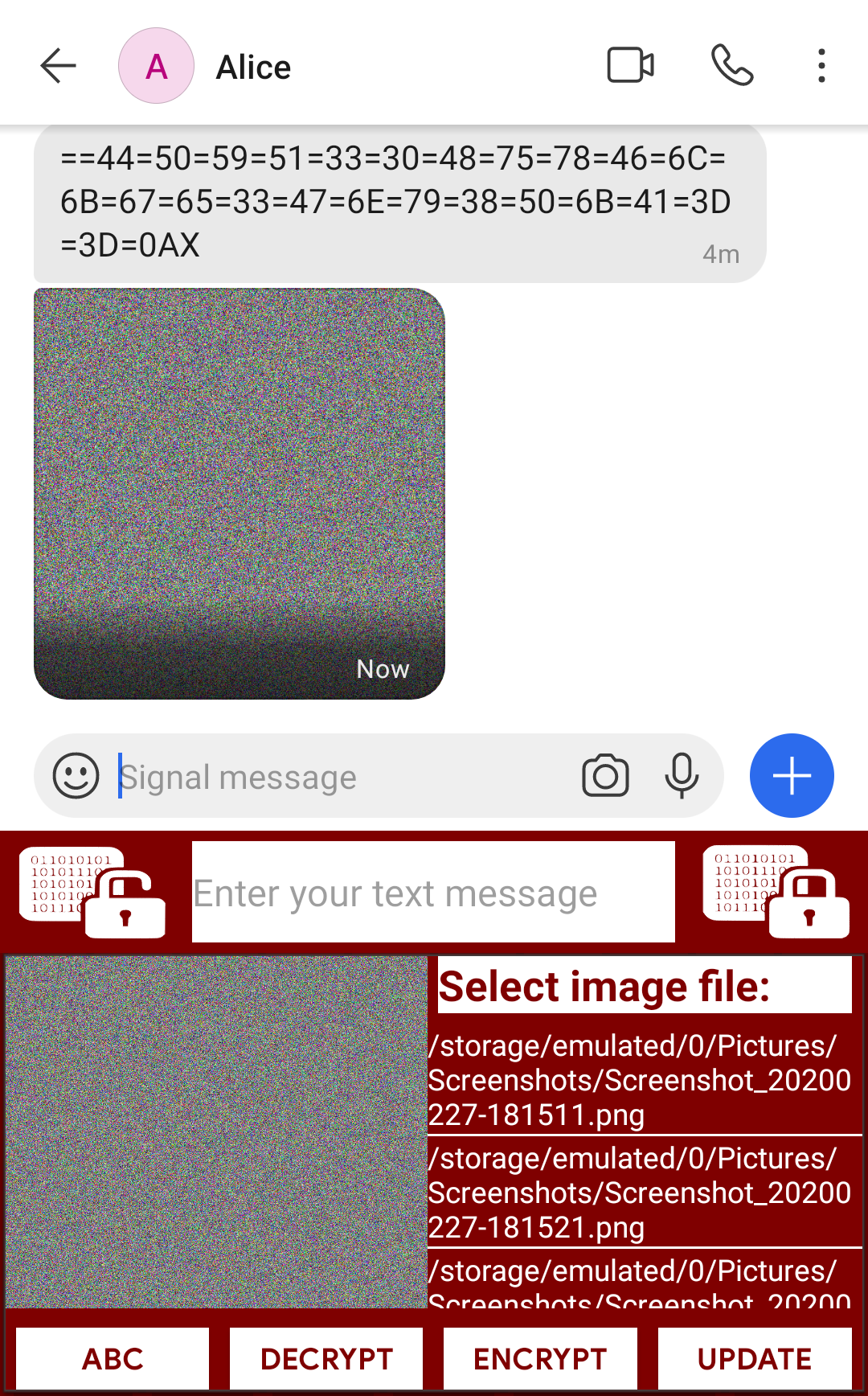}
		\caption{Bob receives the encrypted image file from Alice.}
		\label{fig4:subfig3}
	\end{subfigure}\hspace{3mm}%
	\begin{subfigure}{.22\textwidth}
		\centering
		\includegraphics[width=1\linewidth]{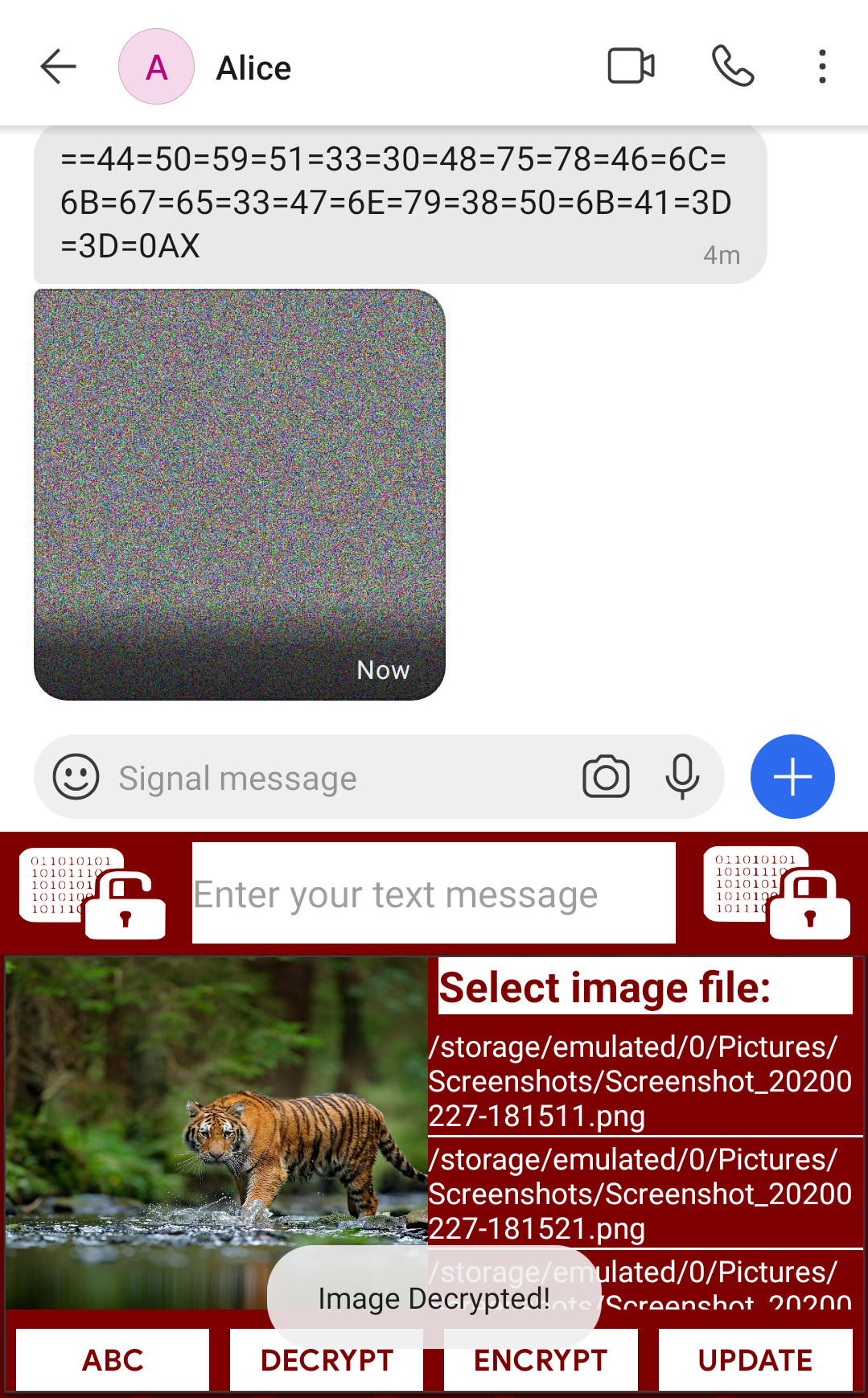}
		\caption{Bob decrypts the encrypted image file locally on his phone device by tapping the DECRYPT button located at the bottom of the encrypted keyboard.}
		\label{fig4:subfig4}
	\end{subfigure}\hspace{3mm}%
	\caption{Encryption and decryption of the image file using our encrypted keyboard when the Signal application is used to send and receive the image file exchanged between Alice and Bob.}
	\label{fig4:user interface encrypt/decrypt image}
	\vspace{-1mm}
\end{figure*}

Not only can our encrypted keyboard encrypt and decrypt text messages, but it can also encrypt and decrypt multimedia messages like images, audio, and video. Figure \ref{fig4:user interface encrypt/decrypt image} shows how our encrypted keyboard is used to encrypt and decrypt an image file when an instant messaging app like Signal is used to send and receive the encrypted image file between two end users. The same steps, as shown in Figure \ref{fig4:user interface encrypt/decrypt image}, can be done if we want to encrypt and decrypt other multimedia elements (such as audio and video) using the additional user interfaces of our encrypted keyboard associated with audio and video tasks (see Figures \ref{fig2:subfig4} and \ref{fig2:subfig5}).

Our encrypted keyboard can encrypt and decrypt multimedia data locally on the user’s phone device. As shown in Figure \ref{fig2:subfig1}, for the purpose of encrypting and decrypting multimedia elements locally on the user’s phone device, the main user interface of our encrypted keyboard has three different keys at the bottom of the interface that allow a user to easily navigate between other interfaces for image, audio, and video tasks. We designed new custom layouts as additional layouts to our main encrypted keyboard layout that allow a user to encrypt and decrypt his multimedia data locally on his phone device (see Figures \ref{fig2:subfig3}, \ref{fig2:subfig4}, and \ref{fig2:subfig5}). Therefore, the user can prevent any IM application from performing the CSS technique on his phone device, thereby securing all his multimedia messages that may be scanned by an IM application. In our encrypted keyboard application, the AES algorithm in CBC mode was used to encrypt and decrypt multimedia data. We used the same key for the encryption and decryption processes of all multimedia messages that were sent between two end users. The user needs to click on the image, audio, or video key button at the bottom of the main user interface of our encrypted keyboard in order to navigate to the related keyboard interface, and therefore can view an exhaustive list of all existing images, audio, or video files on his phone device. Then, from the long list displayed on his current keyboard interface, he can choose an existing image, audio, or video file and encrypt it by clicking the \textit{ENCRYPT} button. After that, the user can send the encrypted file through an IM application (such as Signal) to the intended recipient. Once the intended recipient receives the encrypted file on his phone device through the IM application, he can decrypt the encrypted file by clicking on the \textit{DECRYPT} button, thereby obtaining the original image, audio, or video file that was sent by the sender. The intended recipient then looks at the original image file, listens to the original audio file, or watches the original video file on the encrypted keyboard interface on his phone. We designed on our encrypted keyboard, as shown in Figures \ref{fig2:subfig4} and \ref{fig2:subfig5}, playback control buttons (like play, pause, and stop) in order to allow the user to control the playback of audio and video files. Furthermore, the user will have additional buttons in his audio keyboard layout to record and encrypt voice memos locally on his phone device. As soon as the user finishes recording his voice memo file, the voice memo file will be encrypted immediately, and therefore he can insert it into an IM application to send it to his intended recipient.

\vspace{-1mm}
\section{Keys Management and Distribution}
\label{keysmanagement}
\begin{figure*}[h]
	\centering
	\includegraphics[width=0.85\textwidth]{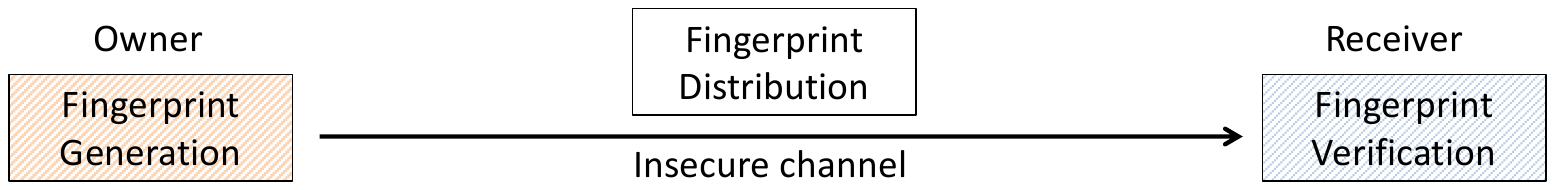}
	\vspace{-63mm}
	\caption{The main components of keys distribution approach. The fingerprint distribution channel is insecure.}
	\label{fig:components}
	\vspace{-1mm}
\end{figure*}

We use a new decentralized approach, which is a public key verification system based on audio fingerprints $R$ consisting of a set of words $W$ spoken in the owner's voice $V$. The basic idea behind this system is for the public key owners to speak the fingerprint of their static/permanent public keys, and for the receivers to authenticate the validity of the public key by verifying the audio fingerprint. To bind the public key to the owners, the audio fingerprint should be verified at two levels: (1) data integrity and (2) voice integrity. To verify that the fingerprint is valid, receivers should verify that the fingerprint matches the hash of the public key (i.e., data integrity) and that the fingerprint is spoken by the owner (i.e., voice integrity). The integrity of data in this system is verified by an \textit{automatic fingerprint comparison} tool that is built on top of speech transcription \cite{shirvanian2017cccp}. This tool automatically converts the audio fingerprint to text and compares it to the hash of the received public key. The attacker who injects his public key should also inject the fingerprint by generating a matching fingerprint in the user's voice. If the attacker only injects the public key but does not change the fingerprint, the receiver can detect the attack. Also, if the attacker speaks his fingerprint, the receiver can detect the attack since the voice speaking the fingerprint does not match the voice of the owner, even though the fingerprint may match the hash of the public key. This system consists of the following components, as depicted in Figure \ref{fig:components}.

\begin{enumerate}[(1)]
	
	\item{\textbf{Fingerprint Generation and Recitation:}}
	Figure \ref{fig:generation} shows the process of generating the fingerprint in this system. Here, the hash of the public key is mapped into PGP words \cite{juola1996whole}, following the approach used by ZRTP \cite{zfoneproject}. The output of the SHA-256 hash is fully mapped to the fingerprint without truncation, resulting in 32 PGP words $W$. The PGP word list consists of two sets of 256 phonetically distinct words, such that they have an \textit{optimum distinction}. To encode a bit string, each byte is mapped into one word from one of the odd and even word sets. Since the even and odd lists are different, human errors in duplicate reading, swapping words, and omission of words are detected. The user generates the audio fingerprint $R$ by speaking and recording $W$ in their own voice characterized by $V$. Since this task takes place only once, the user can spend sufficient effort on preparing acceptably high-quality audio, perhaps using home recording devices in a quiet place. The user can also check (pre-evaluate) the results of the transcription to make sure that $R$ would be perceived well at the receiver's side. Any errors during this phase may be corrected by the user by re-recording the fingerprint.
	
	\begin{figure*}[h]
		\centering
		\includegraphics[width=0.85\textwidth]{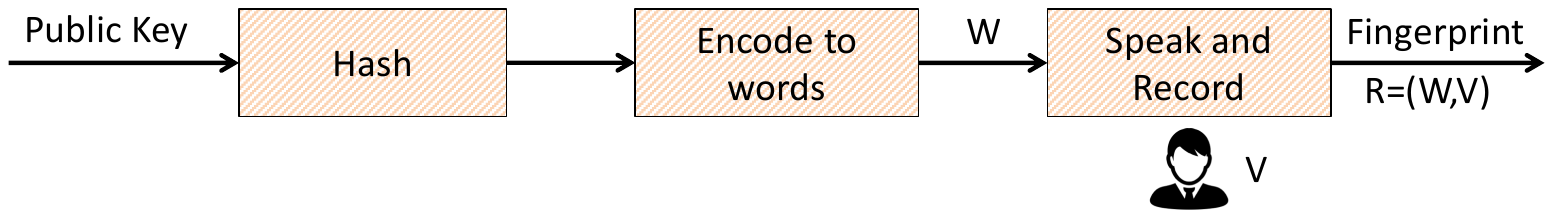}
		\vspace{-63mm}
		\caption{Generating and reciting the audio fingerprint from the public key}
		\label{fig:generation}
		\vspace{-1mm}
	\end{figure*}

	\item{\textbf{Fingerprint Distribution:}}
	This new approach is a distributed public key verification system and does not require imposing trust onto centralized third parties (e.g., a certificate authority) and does not require any trusted auxiliary channel (e.g., out-of-band secure channels). It uses the human voice to authenticate users and binds the public keys to owners' identities. Therefore, in this system, the user can share the fingerprint along with the public key on any public platform accessible by the other parties.

	\item {\textbf{Fingerprint Verification:}}
	Figure \ref{fig:verification} shows the process of verifying the fingerprint in this system. This step represents the core {novel} component of this system, which involves building a fingerprint comparison tool. This tool was built by carefully adapting speaker-independent speech-to-text transcription engines. To bind the public keys to the owners, we rely on the users to verify the public key owners' voice $V$. The users should have prior knowledge of each other's voices.

	\begin{figure*}[h]
		\centering
		\includegraphics[width=0.85\textwidth]{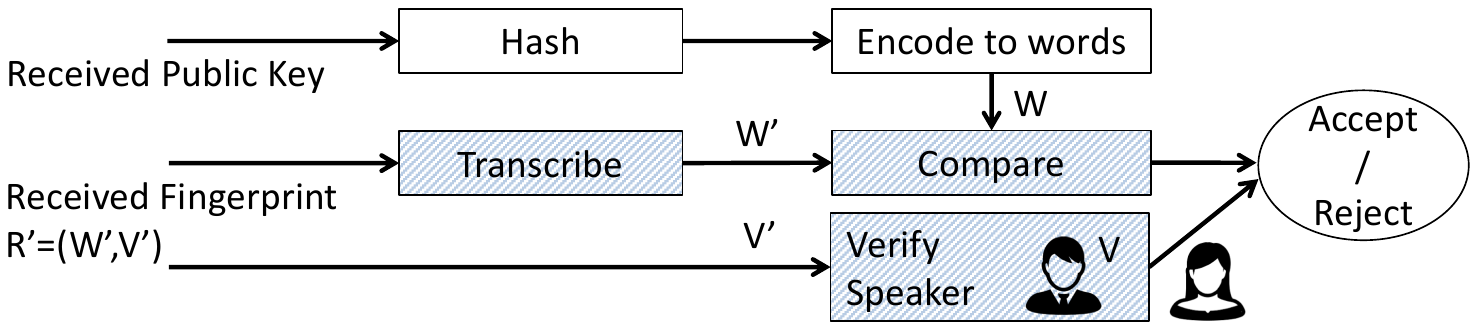}
		\vspace{-55mm}
		\caption{Verifying the received fingerprint}
		\label{fig:verification}
		\vspace{-1mm}
	\end{figure*}
	
\end{enumerate}

This system does not require imposing trust on third parties or a distributed network of trusted users, unlike prior models such as a Public Key Infrastructure (PKI) and web of trust. Each user simply generates a key fingerprint by computing the hash of the public key mapped into PGP words and key owners speak, record, and share the fingerprint with peers (via any out-of-band channel) who can validate the binding between the public key and its owner by: (1) automatically converting the vocalized fingerprint to the textual fingerprint using speech-to-text technology and comparing it with the hash of the public key, and (2) manually verifying that the voice speaking the vocalized fingerprint corresponds to the public key owner. Therefore, the automated vocalized fingerprint comparison tool was built using speech recognition technology. This new system can benefit from any off-the-shelf transcriber (e.g., Google Cloud Speech-to-Text). The fingerprint comparison tool receives the public key and the audio fingerprint $R$ from the owner. The tool generates the hash of the public key and maps it to the words $W$, the same as the fingerprint generation tool. The received audio fingerprint is converted to text $W'$ using the speech-to-text tool and compared with $W$ computed from the received public key. A matching $W$ and $W'$ indicates that the public key is valid, as long as the speaker's voice can be verified by the receivers.

\begin{figure*}[h]
	\centering
	\begin{subfigure}{.22\textwidth}
		\centering
		\includegraphics[width=1\linewidth]{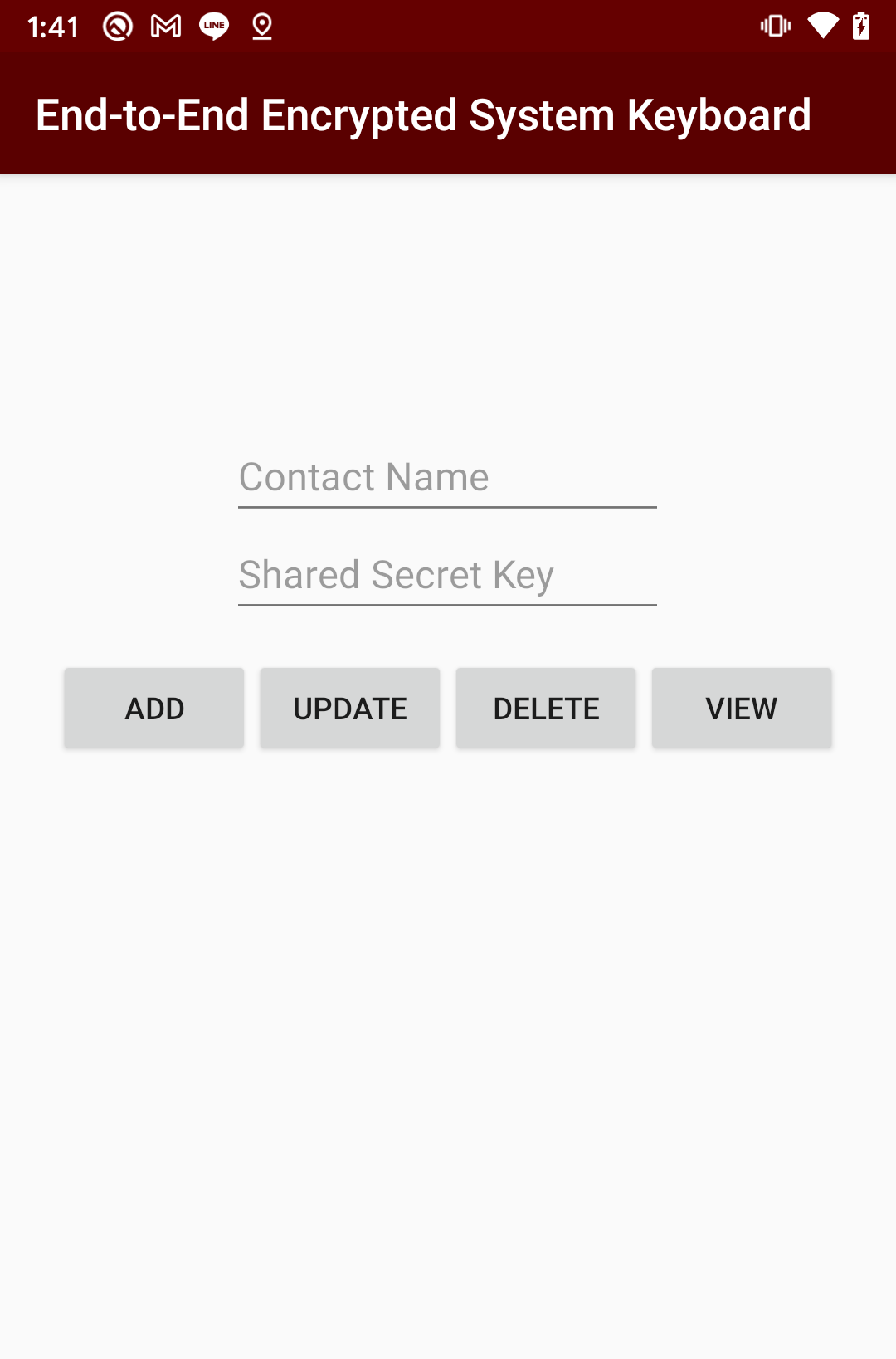}
		\caption{Adding intended recipients and shared secret keys}
		\label{fig5:subfig1}
	\end{subfigure}\hspace{5mm}%
	\begin{subfigure}{.22\textwidth}
		\centering
		\includegraphics[width=1\linewidth]{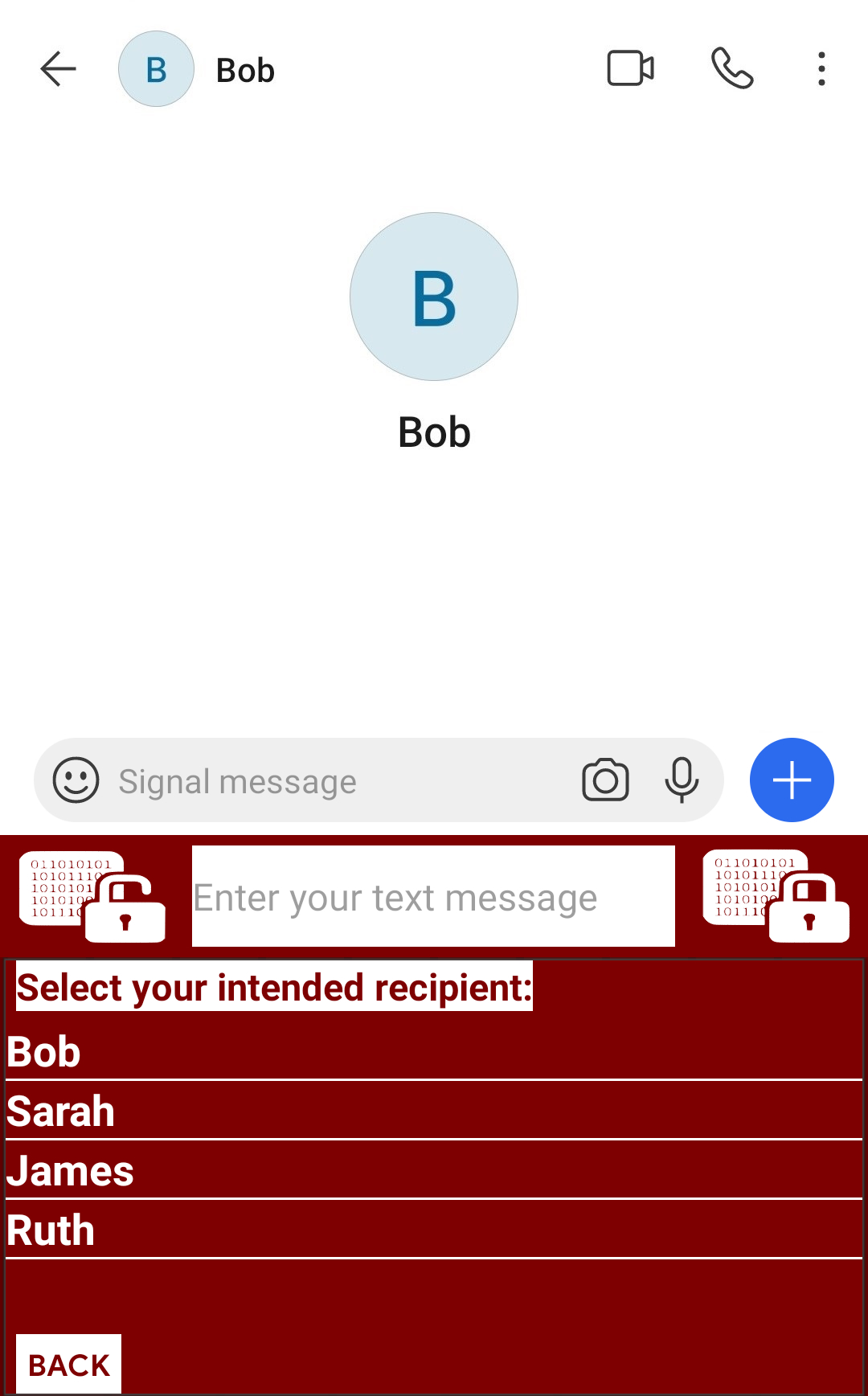}
		\caption{Available intended recipients}
		\label{fig5:subfig2}
	\end{subfigure}\hspace{5mm}%
	\begin{subfigure}{.22\textwidth}
		\centering
		\includegraphics[width=1\linewidth]{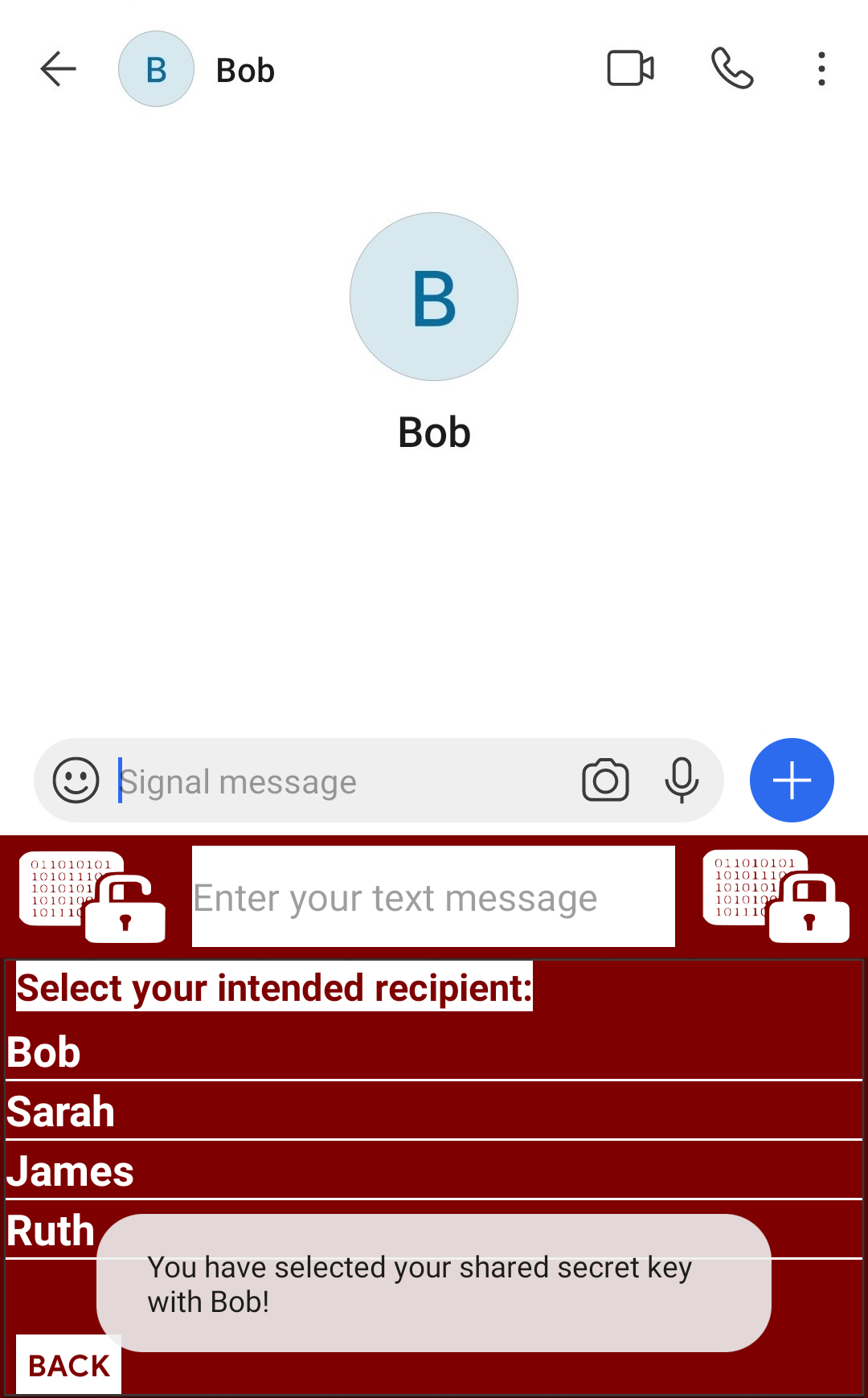}
		\caption{Selecting the shared secret key with Bob}
		\label{fig5:subfig3}
	\end{subfigure}\hspace{5mm}%
	\caption{User interface of using shared secret key}
	\label{fig5:secret key interface}
	\vspace{-1mm}
\end{figure*}

Also, as a part of the fingerprint verification process in this system, receivers should identify the public key owners by their voice, a process that the system refers to as speaker verification. Speaker verification is based on the uniqueness of the sound waves produced by a voice. Since the sound waves reflect the vibrations of air, and such vibrations are dependent on the shape of the oral and nasal cavities above the larynx, each individual's voice should produce a different signal. Besides, each person has a different speech behavioral pattern (e.g., speaking style and accent) that gives each person a unique acoustic feature set, reflecting the body anatomy and speaking style. Thus, the human can identify a particular speaker's voice by recognizing the voice features and behavioral patterns. We assume that users know each other in advance and can recognize each other's voices (the same assumption applies to all other applications that rely on audio fingerprints). Possibly, if they do not have prior knowledge of the speaker's voice, they can make a phone call or listen to publicly available voice samples of the owner (e.g., published on social media). Similar approaches, such as the web of trust, could also be used, in which unknown people would be vouched for by other trusted users. In our scenario, this implies that if a receiver named Bob does not know the voice of a user named Alice, he can rely on some trusted mutual friends or trusted introducers who do know Alice. Thus, users would be able to agree on a shared symmetric key to be used for the encryption/decryption processes of all their textual and multimedia messages. This shared symmetric key can be saved on our encrypted keyboard application and used for encryption and decryption when a user wants to communicate securely with an intended recipient (see Figure \ref{fig5:secret key interface}).

In this system, we assume a model similar to other decentralized trust models for public key verification. In this model, the end users handle the management and distribution of the keys that provide secure communications functionality such as E2EE, decryption, signing, and verification. The end nodes, their devices, and the incorporated tools (including the transcriber, whether performed by a remote service or a localized engine residing on the device) are assumed to be trusted. However, the channel over which the key is distributed may be controlled by the attacker. An attacker can perform a MitM attack to inspect and alter the public key. For example, the attacker can alter an email containing the user's public key to inject an invalid key, or the attacker can compromise the service that holds the public keys and change them as he wishes. 

However, we assume that the attacker does not have full control over the distributed fingerprint. That is, the attacker cannot insert the fingerprint of his public key while speaking it in the user's voice. This assumption comes from the fact that every human has a unique voice. We note that the attacker may be able to generate fake fingerprints in the user's voice by synthesizing or replicating the victim's voice. For example, the attacker may collect the isolated words available in the dictionary in the user's voice and merge them to create a fingerprint in the user's voice. However, if the dictionary consists of rarely used words (or fake words) such as PGP words, the attacker perhaps would not be able to create the synthesized voice, as argued in \cite{bai2016staying}, in contrast to a dictionary of frequently used ones such as digits. Moreover, we will discuss voice authentication and the prevention of voice replication and synthesis attacks in Section \ref{sec:discuss}.

\vspace{-1mm}
\section{Evaluation}
\label{evaluation}

In this section, we will evaluate the effectiveness of our encrypted keyboard application for encrypting and decrypting text and other multimedia elements such as images, audio, and video on various Android phone devices and IM applications. Table \ref{tab1} exhibits the results of our encrypted keyboard evaluation for encrypting and decrypting exchanged messages between two end users. We show the results of only 15 encrypted messages exchanged between two end users using three different phone devices (namely Samsung Android 4, Samsung Android 5, and Google Pixel) and six different IM applications (namely Signal, Viber, Skype, Telegram, WhatsApp, and LINE). We repeated our exchanged messages for encrypting and decrypting text and other multimedia elements, which yielded similar results, in 60 different cases and they are not reported here because of space constraints.

\textbf{Experimental Setup:} We installed our encrypted keyboard application on three different phone devices and selected it to be the default keyboard on all three devices from every phone’s settings. Six different IM applications (i.e., Signal, Viber, Skype, Telegram, WhatsApp, and LINE) were installed on these phone devices to use them for exchanging encrypted textual messages as well as exchanging other encrypted multimedia messages. We created several user accounts on these installed IM applications to utilize them in our experiment. By using these user accounts, each phone device was used either to encrypt textual messages (or other multimedia messages) on one end or to receive the encrypted textual messages (or the encrypted multimedia messages) and decipher them at the other end. We established many communication channels between any two phone devices used in our experiment to evaluate the efficiency of our encrypted keyboard application for encrypting and decrypting textual messages or other multimedia messages. We used our encrypted keyboard enabled on both phone devices to encrypt and decrypt textual messages or other multimedia messages. We also used the installed IM applications on both phone devices to exchange encrypted textual messages (or encrypted multimedia messages) between two end users. We utilized the ADB functionality and the SMASheD platform to push a native service into the \textit{/data/local/tmp/} directory on each phone device. By using that, we are running the native service in the background for taking a screenshot of the contents of a phone device’s screen every second to make sure that the screenshot image is always obtainable and can be used by the OCR engine as an input file. To ensure the perfect performance of the OCR engine on each phone device, a trained data file for the English language was installed simultaneously with the encrypted keyboard application installation on each phone device.

\textbf{Observations:} To measure the effectiveness of our encrypted keyboard application, we can consider the capability of our encrypted keyboard in terms of encrypting a text, image, audio, or video file on one end and deciphering it on the other end. To observe the ability to encrypt and decrypt textual messages using our encrypted keyboard, we typed a text message using our encrypted keyboard and encrypted it on one phone device (the sender). Then, we sent the encrypted textual message via an IM application (i.e., Signal, Viber, or Skype) to the other phone device (the receiver). Using our encrypted keyboard on the receiver’s phone device, we deciphered the encrypted textual message and obtained the plaintext of the original textual message. In addition, to test the effectiveness of our encrypted keyboard application in encrypting and decrypting other multimedia messages, we chose (an image, audio, or video file) and encrypted it on one phone device (the sender). Then, we sent (the encrypted file) over an IM application (i.e., Signal, Viber, Skype, Telegram, WhatsApp, or LINE) to the other phone device (the receiver). By using our encrypted keyboard on the receiver’s phone device, we deciphered (the encrypted file) and therefore obtained (the original image, audio, or video file). Subsequently, we displayed the original image file, played the original audio file, or played the original video file on the receiver’s phone screen. We repeated these processes (encrypting and decrypting text, images, audio, and video files) many times using three different phone devices and six different IM applications (i.e., Signal, Viber, Skype, Telegram, WhatsApp, and LINE). Thus, we observed the capability of our encrypted keyboard to perform both encryption and decryption of textual messages as well as other multimedia messages on these phone devices and IM applications. The results of the sending and receiving of encrypted textual messages as well as other multimedia messages among three different phone devices are shown in Table \ref{tab1}. A total of six different instant messaging applications were utilized to send and receive encrypted textual messages and other multimedia messages. The results show that our encrypted keyboard can encrypt text, images, audio, and video files on one phone device and decipher them on the other phone device. Therefore, our encrypted keyboard application can easily encrypt and decrypt any text, image, audio, or video file that two end users exchange. This protects the user's data from any CSS technique that an IM application could use.

\begin{table*}[t]
	\scriptsize
	\centering
	\caption{Evaluation results of our encrypted keyboard for encrypting and decrypting exchanged messages}\label{tab1}
	\begin{tabular}{|c|c|c|c|c|c|c|c|}
		\hline
		\textbf{\begin{tabular}[c]{@{}c@{}}Message\\ No.\end{tabular}} & \textbf{\begin{tabular}[c]{@{}c@{}}Message\\ Type\end{tabular}} & \textbf{\begin{tabular}[c]{@{}c@{}}Encryption\\ Status\end{tabular}} & \textbf{\begin{tabular}[c]{@{}c@{}}Sender\\ Device\end{tabular}} & \textbf{\begin{tabular}[c]{@{}c@{}}Messaging\\ Application\\ Medium\end{tabular}} & \textbf{\begin{tabular}[c]{@{}c@{}}Receiver\\ Device\end{tabular}} & \textbf{\begin{tabular}[c]{@{}c@{}}Accuracy of\\ OCR Engine\end{tabular}} & \textbf{\begin{tabular}[c]{@{}c@{}}Decryption\\ Status\end{tabular}} \\ \hline
		1                                                              & Text                                                            & Successful                                                           & \begin{tabular}[c]{@{}c@{}}Samsung\\ Android 4\end{tabular}      & Signal                                                                            & \begin{tabular}[c]{@{}c@{}}Samsung\\ Android 5\end{tabular}        & 100\%                                                                     & Successful                                                           \\ \hline
		2                                                              & Text                                                            & Successful                                                           & \begin{tabular}[c]{@{}c@{}}Google\\ Pixel\end{tabular}           & Viber                                                                             & \begin{tabular}[c]{@{}c@{}}Samsung\\ Android 4\end{tabular}        & 100\%                                                                     & Successful                                                           \\ \hline
		3                                                              & Text                                                            & Successful                                                           & \begin{tabular}[c]{@{}c@{}}Samsung\\ Android 5\end{tabular}      & Skype                                                                             & \begin{tabular}[c]{@{}c@{}}Google\\ Pixel\end{tabular}             & 100\%                                                                     & Successful                                                           \\ \hline
		4                                                              & Image                                                           & Successful                                                           & \begin{tabular}[c]{@{}c@{}}Google\\ Pixel\end{tabular}           & Telegram                                                                          & \begin{tabular}[c]{@{}c@{}}Samsung\\ Android 5\end{tabular}        & N/A                                                                     & Successful                                                           \\ \hline
		5                                                              & Image                                                           & Successful                                                           & \begin{tabular}[c]{@{}c@{}}Samsung\\ Android 5\end{tabular}      & WhatsApp                                                                          & \begin{tabular}[c]{@{}c@{}}Samsung\\ Android 4\end{tabular}        & N/A                                                                     & Successful                                                           \\ \hline
		6                                                              & Image                                                           & Successful                                                           & \begin{tabular}[c]{@{}c@{}}Samsung\\ Android 4\end{tabular}      & Signal                                                                            & \begin{tabular}[c]{@{}c@{}}Google\\ Pixel\end{tabular}             & N/A                                                                     & Successful                                                           \\ \hline
		7                                                              & Audio                                                           & Successful                                                           & \begin{tabular}[c]{@{}c@{}}Samsung\\ Android 5\end{tabular}      & LINE                                                                              & \begin{tabular}[c]{@{}c@{}}Samsung\\ Android 4\end{tabular}        & N/A                                                                     & Successful                                                           \\ \hline
		8                                                              & Audio                                                           & Successful                                                           & \begin{tabular}[c]{@{}c@{}}Samsung\\ Android 4\end{tabular}      & Viber                                                                             & \begin{tabular}[c]{@{}c@{}}Google\\ Pixel\end{tabular}             & N/A                                                                     & Successful                                                           \\ \hline
		9                                                              & Audio                                                           & Successful                                                           & \begin{tabular}[c]{@{}c@{}}Google\\ Pixel\end{tabular}           & Skype                                                                             & \begin{tabular}[c]{@{}c@{}}Samsung\\ Android 5\end{tabular}        & N/A                                                                     & Successful                                                           \\ \hline
		10                                                             & \begin{tabular}[c]{@{}c@{}}Voice\\ memo\end{tabular}            & Successful                                                           & \begin{tabular}[c]{@{}c@{}}Samsung\\ Android 4\end{tabular}      & Telegram                                                                          & \begin{tabular}[c]{@{}c@{}}Samsung\\ Android 5\end{tabular}        & N/A                                                                     & Successful                                                           \\ \hline
		11                                                             & \begin{tabular}[c]{@{}c@{}}Voice\\ memo\end{tabular}            & Successful                                                           & \begin{tabular}[c]{@{}c@{}}Google\\ Pixel\end{tabular}           & WhatsApp                                                                          & \begin{tabular}[c]{@{}c@{}}Samsung\\ Android 4\end{tabular}        & N/A                                                                     & Successful                                                           \\ \hline
		12                                                             & \begin{tabular}[c]{@{}c@{}}Voice\\ memo\end{tabular}            & Successful                                                           & \begin{tabular}[c]{@{}c@{}}Samsung\\ Android 5\end{tabular}      & Signal                                                                            & \begin{tabular}[c]{@{}c@{}}Google\\ Pixel\end{tabular}             & N/A                                                                     & Successful                                                           \\ \hline
		13                                                             & Video                                                           & Successful                                                           & \begin{tabular}[c]{@{}c@{}}Google\\ Pixel\end{tabular}           & LINE                                                                              & \begin{tabular}[c]{@{}c@{}}Samsung\\ Android 5\end{tabular}        & N/A                                                                     & Successful                                                           \\ \hline
		14                                                             & Video                                                           & Successful                                                           & \begin{tabular}[c]{@{}c@{}}Samsung\\ Android 5\end{tabular}      & Viber                                                                             & \begin{tabular}[c]{@{}c@{}}Samsung\\ Android 4\end{tabular}        & N/A                                                                     & Successful                                                           \\ \hline
		15                                                             & Video                                                           & Successful                                                           & \begin{tabular}[c]{@{}c@{}}Samsung\\ Android 4\end{tabular}      & Skype                                                                             & \begin{tabular}[c]{@{}c@{}}Google\\ Pixel\end{tabular}             & N/A                                                                     & Successful                                                           \\ \hline
	\end{tabular}
	\vspace{-1mm}
\end{table*}

\vspace{-1mm}
\section{Discussion and Future Work}
\label{sec:discuss}

\subsection{Strengths and Limitations of Our Study}
We believe that our study has several strengths. Our encrypted keyboard covers all users' messages (textual and multimedia messages) just like in day-to-day messaging system use. This is in contrast to current encrypted keyboard applications (such as Enigma Encryption Keyboard and WhisperKeyboard), which only allow the user to encrypt and decrypt textual messages. Also, our encrypted keyboard automates the decryption process of textual messages, whereas, in current encrypted keyboard applications (such as Enigma Encryption Keyboard and WhisperKeyboard), the user needs to copy and paste the text every time he wants to decrypt such a textual message, which may place a very heavy burden on the user. By using the SMASheD server and an OCR mechanism in the decryption process, it helps to automate the text decryption on our encrypted keyboard and, therefore, unburdens the human user from copying and pasting text on the system keyboard. In the decryption process, we also convert the encrypted text into a hexadecimal format to increase the accuracy of the OCR performance. As a rule of thumb, the accuracy of OCR tools can fluctuate every time the OCR engine reads any selected text, recognizing all possible text characters. Therefore, limiting the number of characters to a small group of characters like the hexadecimal symbols will improve the accuracy of OCR performance. As shown in Table \ref{tab1}, we achieved 100\% accuracy on the OCR performance by converting the encrypted text into a hexadecimal format. We believe that recognizing a small group of characters by the OCR engine is better than recognizing all possible text characters if we consider the accuracy of OCR tools, which fluctuates from 71\% to 98\% \cite{patel2012optical}. It helped us reach 100\% accuracy in the OCR performance because we recognized only hexadecimal symbols from 0 to 9, and A to F. To automate our decryption process in textual messaging, our decryption process has to perform two tasks. The first task is to run the OCR engine on a screenshot image to obtain an editable and searchable text. Once the text is extracted from the screenshot image, the second task in our decryption process is performed by reading and extracting an encrypted text to decipher it by using the AES algorithm with the same secret key.

In our study, we focused on studying secure messaging by designing an encrypted keyboard that runs on Android phone devices as a system keyboard. As indicated in Section \ref{evaluation}, we conducted our experiments using only Android phones available in our laboratory. We utilized two older versions and one more recent one. However, we believe that our encrypted keyboard is compatible with every Android phone device, regardless of its version, as it is an Android app that was designed as a system keyboard app. Users can type text on this encrypted keyboard and encrypt it before putting it into an instant messaging application like WhatsApp, Signal, or Viber. Users can then use the IM application to send the encrypted text message to the other intended parties. Likewise, they can use this encrypted keyboard to encrypt other multimedia elements such as images, audio, and video. The main objective of this encrypted keyboard is to prevent any IM application from performing a CSS technique and thereby provide protection for all messages exchanged between end users, including text, images, audio, and video files. By using our encrypted keyboard to encrypt multimedia data before inserting it into instant messaging applications, users can avoid the CSS technique that may be performed by these instant messaging applications when they exchange their messages via these applications. Our encrypted keyboard can also be used to decipher the encrypted multimedia data once the intended recipients receive it in these IM applications. Therefore, our study showed that our encrypted keyboard is a practical and feasible approach that can effectively overcome the CSS technique that could be performed by an IM application (e.g., WhatsApp, Signal, or Viber). However, due to an automated decryption process that relies heavily on a screenshot image of the current phone's screen, we could not have control over the length of the text, especially in long text messages. The phone's screen size (viewport size) is an important factor that needs to be taken into consideration and may affect the process of decrypting encrypted textual messages. Thus, our encrypted keyboard decrypts the encrypted text showing on the current phone's screen, which is based on various screen sizes (viewport sizes) from one phone to another. Further studies should be conducted to cover any length of the text that may exceed the phone's screen size (viewport size) during the automated decryption process of decrypting encrypted textual messages.

\vspace{-1mm}
\subsection{Voice Reordering Attacks}
In \cite{shirvanian2014wiretapping}, a ``voice reordering'' attack was introduced against end-to-end short-spoken text authenticated systems. In the reordering attack, the attacker collects isolated units of the fingerprint (e.g., words or digits) and combines them to create new fingerprints not spoken before. We argue that a reordering attack would be very difficult to pursue if the fingerprint dictionary is sufficiently large. We assume that the dictionary consists of 256 words for even positions and 256 words for odd positions. The easiest attack is to assume that the attacker has obtained ``all the words'' in the dictionary and has built a data set to create ``any desired fingerprint'' in the user's voice by mixing and matching the words from this data set. To collect all the words, we should assume that the user has at least 16 ($=512/32$) unique public keys with no repetitive words in any of them, i.e., the user has 16 public keys mapping to 16 fingerprints $R_i=(W_i,V_i), i \in \{1,...,16\}$ (each $W_i$ is a sequence of 32 words $w_i$, 16 words in even and 16 words in odd positions) such that $ \forall~0 < i,j \leq 32 ,  w_i = w_j \implies i=j $. The probability of the user having 16 different fingerprints with absolutely no repetitive words (in each fingerprint, and among all fingerprints) can be calculated as the following (multiplying the probability of each textual fingerprint $W_i$ not containing any of the dictionary words used in the $i-1$ preceding fingerprints): 


{
	\small{
		\begin{equation}
			\begin{split}
				P1 = \frac{(256-0)^{32}}{256^{32}} \times \frac{(256-16)^{32}}{256^{32}} \times \frac{(256-32)^{32}}{256^{32}} .... \\ \times \frac{(256-240)^{32}}{256^{32}} = 
				\frac{15!}{{16}^{16}} = 7.09E-8
			\end{split}
	\end{equation}}
}

As can be seen, the probability of the attacker collecting all the words in the dictionary (even if we generously assume that the user has 16 different public keys) is very low.

On the other end of this statistic, we can assume that the user has only 1 fingerprint (i.e., 1 public key). Therefore, the attacker has collected a total of 16 words (assuming that the fingerprint does not contain any duplicate words) from each of the even and odd sets. The attacker tries to create a public key such that the words in the attacker's fingerprint match the words used in the user's fingerprint (but perhaps not in the same order). The probability of succeeding in this attack is: 


{
\small{
	\begin{equation}
		P2 = \frac{{16}^{16}}{{256}^{32}} = \frac{1}{{16}^{16}} = 5.42E-20
\end{equation}}
}

Note that in any other situation (i.e., the user having more than 1 and less than 16 unique fingerprints), the attacker has access to only a subset of the dictionary in the user's voice. Therefore, $P3$, which is the probability of creating a fingerprint containing only the spoken words, falls between $P1$ and $P2$, i.e., $P1 < P3 < P2$. Note that increasing the size of the dictionary can also help to reduce the chance of a reordering attack. Finally, if the attacker creates a fingerprint by mixing some of the words from the spoken data set and replacing the rest with synthesized voices, such inconsistency and the use of synthesized voices should be detected (since we expect the user to listen to the audio samples and verify the speaker).

\vspace{-1mm}
\subsection{Attacks on Speech Recognition}
A few types of attacks against transcription technology have been proposed recently \cite{carlini2016hidden,vaidya2015cocaine}. These attacks generate \textit{audible} samples that are not intelligible to the human user but interpretable by the transcriber. Although these types of attacks may be used against virtual personal assistant applications to run the attacker's commands, we assume that in the public key verification system, the user who listens to the vocalized fingerprint to verify the speaker can detect such suspiciously malformed audible/robotic audio samples and would notice if the content of the vocalized fingerprint is completely different from the fingerprint (e.g., music being played in the background). Inaudible attacks on speech transcription, such as the one proposed in \cite{zhang2017dolphinattack}, require physical access to the speech transcription device. Besides, such an approach should have knowledge of the hardware characteristics of the speech transcription device and therefore is not relevant to this system and cannot compromise the security of online transcription systems.

\vspace{-1mm}
\section{Conclusions}
In this paper, we introduce an encrypted keyboard built as a system keyboard. Our encrypted keyboard can be enabled by users on their phone devices as a default system keyboard in order to use it on every application that prompts users to enter some data. Our encrypted keyboard can offer protection against CSS technologies that might be implemented by a vast number of IM applications. Besides the protection against CSS technologies, our encrypted keyboard can be used to strengthen security against MitM attackers by adding an extra E2EE layer on top of the current E2EE functionality implemented by many end-to-end encrypted applications. Users can use our encrypted keyboard to encrypt their multimedia data locally on their phone devices before exchanging it over an IM application. They can also use our encrypted keyboard to decipher all encrypted multimedia data received from an IM application locally on their phone devices. Our work shows that our encrypted keyboard can successfully encrypt and decrypt all sending and receiving messages through IM applications. Therefore, our encrypted keyboard can prevent any CSS system that might be implemented by an IM application. It can also be used to reinforce the security of E2EE functionality against MitM attacks by providing another E2EE functionality in addition to the current E2EE feature provided by end-to-end encrypted applications. Therefore, users can have a duplicate encryption scheme when they use our encrypted keyboard to encrypt their messages and then exchange them via end-to-end encrypted applications. This duplicate encryption scheme gives our encrypted keyboard a strong defense mechanism against MitM attackers, even if they can somehow compromise the E2EE functionality provided by end-to-end encrypted applications.

\subsubsection{Acknowledgment.}
The authors would like to thank Maliheh Shirvanian who contributed to the audio fingerprinting approach.

%
%
%

\bibliographystyle{splncs04}
\bibliography{encKeyboard_references}

\end{document}